\documentclass[%
 aip,
rsi,%
 amsmath,amssymb,
 reprint,%
]{revtex4-1}

\usepackage[dvipsnames]{xcolor}
\usepackage{graphicx}% Include figure files
\usepackage{dcolumn}% Align table columns on decimal point
\usepackage{bm}
\usepackage{lipsum}

\usepackage{dcolumn}% Align table columns on decimal point
\usepackage{epstopdf}
\usepackage{xcolor}
\usepackage{natbib}
\usepackage{subcaption} 
\usepackage{float}
\usepackage{tabularx}
\usepackage{mathrsfs}
\usepackage{upgreek}% non-italic greek symbols
\usepackage[a]{esvect}
\setcitestyle{square}
\usepackage[toc,page]{appendix}
\usepackage[font={small,normalfont}]{caption}
\usepackage[colorlinks=true, allcolors=blue]{hyperref}
\usepackage[normalem]{ulem}
\usepackage[capitalise]{cleveref}

\newcommand{\rc}[1]{\textcolor{black}{#1}}

\begin{document}

\title{Enhanced energy gain through higher-order resonances during direct laser acceleration with superluminal phase velocity}

\author{I-L. Yeh}
\email[]{iyeh@ucsd.edu}

\affiliation{Department of Physics, University of California San Diego, La Jolla, CA 92093}

\author{K. Tangtartharakul}
\affiliation{Department of Mechanical and Aerospace Engineering, University of California San Diego, La Jolla, CA 92093}

\author{R. Bhakta}
\affiliation{Department of Mechanical and Aerospace Engineering, University of California San Diego, La Jolla, CA 92093}

\author{H. Tang}
\affiliation{Gérard Mourou Center for Ultrafast Optical Science, University of Michigan, Ann Arbor, Michigan 48109}

\author{L. Willingale}
\affiliation{Gérard Mourou Center for Ultrafast Optical Science, University of Michigan, Ann Arbor, Michigan 48109}

\author{A. Arefiev}
\email[]{aarefiev@ucsd.edu}

\affiliation{Department of Mechanical and Aerospace Engineering, University of California San Diego, La Jolla, CA 92093}

\affiliation{Center for Energy Research, University of California San Diego, La Jolla, CA 92093}

\date{\today}

\begin{abstract}
Ultra-high intensity laser-plasma interactions can produce ultra-relativistic electrons via direct laser acceleration, assisted by quasi-static plasma magnetic and electric fields. These fields transversely confine electron motion and induce betatron oscillations. The net energy gain is strongly influenced by the interplay between two frequencies: the betatron frequency and the frequency of laser field oscillations experienced by the electron. Prior work has shown that energy gain is enabled by a resonance between the betatron oscillations and the oscillations of the laser field. In particular, higher-order resonances occur when the laser field completes multiple cycles during one betatron oscillation, allowing additional regimes of energy transfer beyond the fundamental (betatron) resonance. In this work, we demonstrate that such resonances become particularly effective when the laser’s phase velocity is superluminal. Although the two frequencies generally evolve differently with increasing electron energy — leading to detuning — a superluminal phase velocity introduces a non-monotonic frequency ratio with a global minimum. This minimum allows sustained frequency matching over a broad energy range, thereby enabling enhanced energy gain. As the phase velocity increases, the betatron resonance becomes ineffective due to premature frequency detuning. At the same time, higher-order resonances become increasingly effective, emerging as the dominant mechanisms for enhanced energy gain in this regime of direct laser acceleration.
\end{abstract}

\maketitle

%*******************************************************

\section{Introduction} \label{sec: intro}

In ultra-high-intensity laser-plasma interactions, direct laser acceleration (DLA)~\cite{pukhov1999DLA,arefiev2016beyond,Robinson_2013_PRL} is a process that facilitates energy transfer from the laser beam to plasma electrons. DLA has numerous applications, including generating secondary particle beams such as ions \cite{dover.prl.2020, bailly.pre.2020}, positrons \cite{cowan_positron, Chen_2009_PRL, Chen_2010_PRL, he.njp.2021, he.commphys.2021,sugimoto2023positronPRL,martinez.pre.2025}, neutrons \cite{Pomerantz-PhysRevLett.113.184801, guenther.natcomm.2022}, and producing bright x-ray/gamma-ray sources \cite{Stark2016PRL,wang_power_PRA2020,shen.apl.2021,guenther.natcomm.2022,yeh.pop.2024, Tangtartharakul.NJP.2025}. Recently, DLA has gained more interest due to the construction of petawatt-scale and multi-petawatt laser facilities like ELI \cite{weber2017p3,gales2018eli-np,lureau.hplse.2020, cernaianu.mre.2025}, ZEUS \cite{Nees.zeus.2020,willingale.zeus.2023}, BELLA \cite{nakamura.bella.2017}, and CoReLS \cite{yoon.optica.2021}. 

Research on DLA assisted by quasi-static plasma electric and magnetic fields is particularly active, utilizing analytical theory \cite{Khudik-POP_2016,arefiev.pre.2020,Wang.pop.2020}, numerical simulations \cite{pukhov1999DLA,gong.pre.2020,kemp.pop.2020,babjak.njp.2024,babjak.prl.2024,valenta.pre.2024}, and experimental studies \cite{peebles.njp.2017,Willingale_2018,rosmej.njp.2019,rosmej.ppcf.2020, hussein.njp.2021,tang.njp.2024,cohen.sciadv.2024}. These quasi-static fields, typically radial electric and azimuthal magnetic plasma fields, result from laser propagation through the plasma. They have been consistently observed in particle-in-cell (PIC) simulations~\cite{pukhov1999DLA, arefiev2016beyond, gong.pre.2020} and experiments~\cite{hussein.njp.2021}.

At relativistic laser intensities, DLA generates forward-moving relativistic electrons that are suitable for applications. The electron energy gain in this process is of primary importance. Electrons typically gain their energy, \(\varepsilon_e\), from the transverse electric field of the laser, \(E_{\perp}\). The rate of energy increase is given by \(d \varepsilon_e / dt = - |e| E_{\perp}(t) v_{\perp}(t)\), where \(e\) is the electron charge and \(v_{\perp}\) is the transverse electron velocity. The inherent oscillations of \(E_{\perp}(t)\) and \(v_{\perp}(t)\) make achieving prolonged energy gain far from straightforward.

Electron dephasing is one of the key factors limiting electron energy gain. Electrons inevitably experience dephasing or slippage relative to the laser wavefronts because the phase velocity, \(v_{ph}\), inside a plasma is superluminal (\(v_{ph} > c\)), where \(c\) is the speed of light. At the electron's location, the laser field \(E_{\perp}\) oscillates with a frequency \(\omega'\) that can be much lower than the laser frequency \(\omega_0\) due to the ultra-relativistic forward motion of the electron. The timescale to slip by one wavefront is thus set by \(1/\omega'\). In the absence of plasma fields, this dephasing terminates the electron energy gain, limiting it to no more than one-quarter of the laser field oscillation at the electron's location~\cite{arefiev2016beyond}.

Quasi-static plasma fields can mitigate the negative effects of dephasing. Rather than slowing down the dephasing, this mitigation is typically achieved by inducing transverse electron deflections, thereby altering the direction of the transverse electron velocity \(v_{\perp}\). The characteristic frequency of these deflections, known as the betatron frequency (\(\omega_{\beta}\)), represents the frequency of transverse oscillations induced by the plasma fields. If $\omega_{\beta}$ matches $\langle \omega' \rangle$, the average of \( \omega'\) over one full betatron oscillation, then the transverse velocity can remain anti-parallel to the laser electric field for an extended period of time,  even as dephasing causes oscillations in the field at the electron’s location. This alignment allows the resulting energy gain to significantly exceed the energy gain observed in the absence of plasma fields~\cite{arefiev2016beyond}.

The frequency matching described by \(\omega_{\beta} \approx \langle \omega' \rangle\) is commonly known as betatron resonance~\cite{pukhov1999DLA}. However, this is not a conventional linear resonance because both \(\omega_{\beta}\) and \( \langle \omega' \rangle\) depend on the electron's energy. These dependencies differ, leading to frequency detuning that ultimately limits energy gain. The betatron resonance, or first-order resonance, has been extensively studied and is considered the primary mechanism for energy gain during DLA~\cite{pukhov1999DLA, gong.pre.2020, jirka.njp.2020}.

In addition to the first-order resonance, there are also higher-order resonances defined by \( \langle \omega' \rangle \approx \ell \omega_{\beta}\), where \(\ell > 1\) is an odd integer~\cite{Khudik-POP_2016}. We use the term resonance to indicate that specific frequency matching is required, but, in higher-order resonances, energy gain occurs solely due to the modulation of \(\omega'\) induced by the betatron oscillations~\cite{arefiev.pop.2024}. Unlike the betatron resonance, there is no net energy gain per betatron oscillation in the absence of modulation, as the energy gain is entirely offset by energy loss. The concept of higher-order resonances in the context of DLA was first introduced in Ref.~[\onlinecite{Khudik-POP_2016}]. The effect of the electron propagation angle was subsequently examined in Ref.~[\onlinecite{li.prab.2021}]. Similar to the betatron resonance, higher-order resonances are also influenced by frequency detuning, which limits electron energy gain. Therefore, their effective utilization depends on identifying regimes where detuning can be minimized, thereby enhancing energy gain.

In this paper, we examine how the frequency ratio \( \langle \omega' \rangle / \omega_{\beta} \) evolves with electron energy. We use a test-electron model with prescribed laser and plasma fields to determine the electron dynamics, energy gain, and frequency ratio. The results reveal a counterintuitive benefit of a superluminal phase velocity: it produces a non-monotonic dependence of the frequency ratio on \( \gamma \), as first noted in Ref.~[\onlinecite{Khudik-POP_2016}]. The resulting global minimum allows the electron to remain in resonance as it gains energy. We show that both the betatron resonance and higher-order resonances can benefit from this feature. A key finding is that the optimal phase velocity depends on the resonance order. As a result, higher-order resonances provide a viable path to enhanced energy gain in regimes where the phase velocity is too high for the betatron resonance to remain effective.

The remainder of the paper is organized as follows. In \cref{sec: test-model}, we introduce the test-electron model used to investigate electron dynamics in prescribed laser and plasma fields. In \cref{sec: frequencies}, we define two key frequencies that govern energy exchange and derive an analytical expression for their ratio, which serves as a central diagnostic throughout the paper. In \cref{sec: 1stres} and \cref{sec: hres}, we examine representative examples showing how superluminal phase velocity modifies electron motion in resonance with the laser field, with a focus on the betatron and third-order resonances. In \cref{sec: scan}, we perform a broad parameter scan to identify the conditions under which different resonances produce enhanced energy gain. Finally, in \cref{sec: summary}, we summarize the main findings and discuss their implications.

%%%%%%%%%%%%%%%%%%%%%%%%%%%%%%%%%%%
%%%%%%%%%%%%%%%%%%%%%%%%%%%%%%%%%%%
%%%%%%%%%%%%%%%%%%%%%%%%%%%%%%%%%%

\section{Test-electron model: recap} \label{sec: test-model}

In this paper, we analyze electron dynamics using a test-electron model where the laser and plasma fields are prescribed rather than computed self-consistently. This approach is particularly well-suited for scenarios like DLA, where the energetic electrons under study are a minority. Various test-electron models have been employed to analyze DLA, primarily differing in their approximations of laser and plasma fields \cite{pukhov1999DLA,arefiev2016beyond,Khudik-POP_2016,arefiev.pre.2020,gong.pre.2020,kemp.pop.2020,babjak.prl.2024,valenta.pre.2024}. Our model follows the approach described in Ref.~[\onlinecite{arefiev.pop.2024}], where a more comprehensive discussion can also be found.

%Our model follows the approach described in Ref.~[\onlinecite{arefiev.pop.2024}], and we provide only a brief overview here. For a more comprehensive discussion, readers are encouraged to refer to Ref.~[\onlinecite{arefiev.pop.2024}].

In our model, the time evolution of the electron momentum $\bm{p}$ and position $\bm{r}$ are described by the following system:
\begin{eqnarray}
&& \frac{d \bm{p}}{d t} = - |e| \bm{E} - \frac{|e|}{\gamma m_e c} \left[ \bm{p} \times \bm{B} \right], \label{dpdt} \\
&& \frac{d \bm{r}}{d t} = \frac{\bm{p}}{\gamma m_e c}, \label{drdt} 
\end{eqnarray}
where $m_e$ is the electron mass and 
\begin{equation} \label{gamma}
    \gamma = \sqrt{1 + \bm{p}^2/ m_e^2 c^2}
\end{equation}
is the relativistic factor of the electron. The electric and magnetic fields ($\bm{E}$ and $\bm{B}$) are a superposition of oscillating laser fields (denoted by the subscript ``laser'') and static plasma fields (denoted by the subscript ``stat''): $\bm{E} = \bm{E}_{\rm{laser}} + \bm{E}_{\rm{stat}}$ and $\bm{B} = \bm{B}_{\rm{laser}} + \bm{B}_{\rm{stat}}$.

We approximate the laser by a plane linearly polarized electromagnetic wave with a phase velocity $v_{ph}$ and frequency $\omega_0$  propagating in the positive direction along the $x$-axis. The laser fields are
\begin{eqnarray}
     &&\bm{E}_{\rm{laser}} = \bm{e}_y E_0 \cos(\xi), \label{E-laser}\\
     &&\bm{B}_{\rm{laser}} = \bm{e}_z E_{\rm{laser}} / u, \label{B-laser}
\end{eqnarray}
where $E_0$ is a constant laser amplitude,  
\begin{equation}
    u \equiv v_{ph} / c
\end{equation}
is the normalized phase velocity, 
\begin{equation} \label{xi}
    \xi = \omega_0 t - \omega_0 x/v_{ph} + \xi_* = \omega_0 t - \omega_0 x/ u c + \xi_*
\end{equation}
is the laser phase, with $\xi_*$ being the initial phase at the initial electron location ($x=0$) at $t=0$. The vacuum laser wavelength for this wave is
\begin{equation} \label{lambda_0}
    \lambda_0 = 2 \pi c / \omega_0.
\end{equation}
It is convenient to characterize the field strength using a normalized field amplitude, defined as
\begin{equation}
    a_0 \equiv |e| E_0 / m_e c \omega_0.
\end{equation}

Previous studies using PIC simulations~\cite{pukhov1999DLA, arefiev2016beyond, gong.pre.2020} have shown that a laser beam can generate a channel in the plasma with slowly varying radial electric and azimuthal magnetic fields. We adopt this setup for our model, treating the plasma fields as static. In our study, we focus on flat particle trajectories in the \((x,y)\)-plane (\(z = 0\) and \(p_z = 0\)), intersecting the axis of the channel at \(z = 0\). We thus only need the knowledge of $\bm{B}_{\rm{stat}}$ and $\bm{E}_{\rm{stat}}$ in the $(x,y)$-plane. The radial electric field in the $(x,y)$-plane is
\begin{equation}
    \bm{E}_{\rm{stat}} = 2 \kappa \frac{y}{\lambda_0^2} \frac{m_e c^2}{|e|} \bm{e}_y  \label{E_stat}
\end{equation}
where 
\begin{equation} \label{eq: def kappa}
    \kappa \equiv \pi^2 \omega_{ch}^2 / \omega_0^2
\end{equation} is a dimensionless parameter quantifying the charge density $\rho_{ch}$ that sustains the field, with 
\begin{equation}
    \omega_{ch}^2 \equiv 4 \pi |e| \rho_{ch}/m_e
\end{equation}
being the plasma frequency calculated using $\rho_{ch}$. The plasma magnetic field in the \((x,y)\)-plane, sustained by a current density of \(j_x = -j_0\) (where \(j_0 > 0\)), is given by 
\begin{equation}
    \bm{B}_{\rm{stat}} = - \frac{m_e c^2}{|e|} \frac{2 \alpha y}{\lambda_0^2} \bm{e}_z,
    \label{B_stat}
\end{equation}
where the dimensionless parameter 
\begin{equation} \label{eq: def alpha}
    \alpha \equiv \pi \lambda_0^2 j_0/J_A
\end{equation}
represents the ratio of the current flowing through a circular area with radius \(\lambda_0\) to the classical Alfvén current \(J_A = m_e c^3 / |e|\).

In summary, our test-electron model consists of Eqs.~(\ref{dpdt}) and (\ref{drdt}), where the fields are the laser fields given by Eqs.~(\ref{E-laser}) and (\ref{B-laser}) and the plasma fields given by Eqs.~(\ref{E_stat}) and (\ref{B_stat}). In addition to initial conditions, the model requires four dimensionless  input parameters: $a_0$, $\alpha$, $\kappa$, and $u$. 

To isolate the effect of $u$ on  electron dynamics, we fix the remaining parameters by assigning specific values used consistently throughout this work. Specifically, we set $a_0 = 10$, $\alpha = 1.52$, and $\kappa =  0.44$ for all numerical calculations involving the test-electron model. This parameter choice is informed by particle-in-cell simulations using a setup similar to that described in Ref. [\onlinecite{gong.pre.2020}]. In the numerical examples shown  in \cref{sec: 1stres} and \cref{sec: hres}, the initial phase is set to $\xi_* = \pi/2$. In \cref{sec: scan}, we consider several different values of $\xi_*$ when performing a broad parameter scan. 

%++++++++++++++++++++++++++++++++++++++++

\section{Analysis of frequencies determining electron energy gain} \label{sec: frequencies}

In this section, we derive analytical expressions for two key frequencies that influence electron energy gain from the laser in the presence of static plasma fields. Instead of developing a general solution, we focus on analyzing regimes where the net energy gain by the electron significantly exceeds its oscillatory energy in the laser field. In these regimes, the electron undergoes multiple betatron oscillations across the channel while gaining energy from the laser. This implies that the electron's \(\gamma\)-factor varies slowly on the time scale of a single betatron oscillation. This feature allows us to simplify the analysis of an otherwise complex dynamical system. Our approach builds on the methodology first introduced in Ref.~[\onlinecite{Khudik-POP_2016}] and subsequently expanded in Ref.~[\onlinecite{arefiev.pop.2024}]. We use the same notations as in Ref.~[\onlinecite{arefiev.pop.2024}] to facilitate easy cross-referencing for readers interested in a more detailed derivation.

Our primary focus is on the net electron energy gain during DLA, so it is appropriate to begin by examining how the electron exchanges energy with the fields in the system under consideration. The general energy balance equation, derived from Eq.~(\ref{dpdt}) by taking the scalar product with \(\bm{p}\), is given by:
\begin{equation}
    \frac{d}{dt} \left[ \gamma m_e c^2 \right] = - \frac{|e| (\bm{E} \cdot \bm{p})}{\gamma m_e},
\end{equation}
where the left-hand side represents the rate of change in the electron's kinetic energy, and the right-hand side represents the energy exchange with the electric field \(\bm{E}\) acting on the electron. In our model, \(\bm{E} = \bm{E}_{\rm{laser}} + \bm{E}_{\rm{stat}}\). We utilize the fact that \(\bm{E}_{\rm{stat}}\), given by \cref{E_stat}, is static and express its contribution as a time derivative. Furthermore, since the laser field has only a transverse component, the \(y\)-component given by Eq.~(\ref{E-laser}), we obtain:
\begin{equation} \label{energy exchange 1}
    \frac{d}{dt} \left[ \gamma  + \kappa \frac{y^2}{\lambda_0^2}  \right] m_e c^2 = - |e| E_0 v_y(t) \cos(\xi).
\end{equation}

To determine the net energy gain, we integrate \cref{energy exchange 1} over a single betatron oscillation, effectively removing the energy fluctuations caused by the static plasma electric field. Consider an oscillation that begins at \(y = 0\) at time \(t = t_0\) and has a period \(\Delta t\). After one complete betatron oscillation, the change in the electron's relativistic factor $\gamma$ is given by:
\begin{equation} \label{energy exchange 2}
    \Delta \gamma = - \frac{|e| E_0}{m_e c^2} \int_{t_0}^{t_0 + \Delta t} v_y(t) \cos(\xi) dt.
\end{equation}
In general, the functions under the integral, \(v_y\) and \(\cos(\xi)\), oscillate at different frequencies. The interaction between these oscillations can significantly reduce \(\Delta \gamma\). In the regime of interest, the laser field does not substantially alter \(v_y\), so the electron's velocity oscillates with the betatron frequency, \(\omega_{\beta}\). The laser field itself oscillates with a frequency \(\omega'\), which we refer to as the laser field's oscillation frequency at the electron's location. It is influenced by the electron's motion and differs from \(\omega_0\). \Cref{energy exchange 2}  suggests that \(\Delta \gamma\) can be maximized by appropriately matching \(\omega_{\beta}\) and \(\omega'\). 

Achieving and maintaining a match between \(\omega_{\beta}\) and \(\omega'\) as the electron gains energy requires an understanding of how these frequencies scale with \(\gamma\). In the remainder of this section, we derive the relevant expressions under the assumption that the electron is already ultrarelativistic. We also assume that the electron is moving forward, with 
\begin{equation} \label{cond: 1}
    p_x \gg |p_y| \gg m_e c.
\end{equation}
Additionally, we assume that
\begin{equation} \label{delta u def}
    \delta u \equiv (v_{ph} - c)/c = u - 1 \ll 1.
\end{equation}
This is a reasonable assumption, as PIC simulations indicate that the stable laser propagation necessary for DLA typically occurs under conditions where \(\delta u \ll 1\).

To derive \(\omega_{\beta}\), we first need to determine the amplitude of the betatron oscillations, \(y_*\). This can be achieved by employing the following expression:
\begin{equation} \label{S_exp}
   S =  \gamma - u \frac{p_x}{m_e c} + \left[ u \alpha + \kappa \right]  \frac{y^2}{\lambda_0^2}.
\end{equation}
This expression is a constant of motion in our model, as can be verified using Eqs.~(\ref{dpdt}) and (\ref{drdt}). Given the conditions specified by \cref{cond: 1} and \cref{delta u def}, we have \(u p_x/m_e c \approx p_x/m_e c + \delta u \gamma \), where \(p_x/m_e c\) in the second term on the right-hand side has been approximated by \(\gamma\). Substituting this expression into \cref{S_exp} and rearranging the terms yields:
\begin{equation}
    \gamma - \frac{p_x}{m_e c} \approx S + \gamma \delta u  -  \left[ u \alpha + \kappa \right] \frac{y^2}{\lambda_0^2}.
    \label{S_exp-3}
\end{equation}
As the electron moves away from the axis and \( |p_y|/m_e c \) decreases, the value of the expression on the left-hand side drops. Since this value must remain positive, we can find the maximum transverse displacement, $y_*$, by finding the value of \(y^2\) that causes the right-hand side to go to zero. It follows from \cref{S_exp-3} that \( |y| \leq y_* \), where
\begin{equation}
\label{ymax}
    \frac{y_*}{\lambda_0} = \left[ \frac{S + \gamma \delta u}{\kappa + u \alpha} \right]^{1/2}.
\end{equation}
This expression indicates that, as the electron gains energy, the amplitude of its betatron oscillations increases.

We define the frequency of betatron oscillations as
\begin{equation} \label{def: omega beta}
    \omega_{\beta} \equiv d\psi / dt,
\end{equation}
where \(\psi\) represents the betatron phase, implicitly related to the amplitude \(y_*\) by the expression:
\begin{equation} \label{y beta 0}
    y = y_* \sin \psi.
\end{equation}
Substituting this expression for \(y\) into \cref{S_exp-3} and using the expression for \(y_*\) from \cref{ymax}, we obtain:
\begin{equation}
    \gamma - \frac{p_x}{m_e c} \approx (S + \gamma \delta u) \cos^2 \psi.
    \label{S_exp-4}
\end{equation}
Meanwhile, under the condition given by \cref{cond: 1}, we have:
\begin{equation}
    \gamma - \frac{p_x}{m_e c} \approx \frac{p_x}{m_e c} \frac{p_y^2}{2 p_x^2} \approx \frac{\gamma}{2} \frac{v_y^2}{c^2}.
    \label{S_exp-5}
\end{equation}
Here, we replaced \(p_x/m_e c\) with \(\gamma\) and then used the relation \(p_y = \gamma m_e v_y\) from Eq.~(\ref{dpdt}) to express the result in terms of \(v_y\). Taking the time derivative of \cref{y beta 0}, we find 
\begin{equation} \label{v_y}
    v_y = y_* \omega_{\beta} \cos \psi.
\end{equation}
Substituting this into the right-hand side of \cref{S_exp-5} yields:
\begin{equation}
    \gamma - \frac{p_x}{m_e c} \approx \frac{\gamma}{2} \frac{y_*^2 \omega_{\beta}^2}{c^2} \cos^2 \psi.
    \label{S_exp-6}
\end{equation}
To ensure consistency with \cref{S_exp-4}, the betatron frequency \(\omega_{\beta}\) must satisfy:
\begin{equation} \label{eq: omega beta 0}
    \frac{\omega_{\beta}}{\omega_0} =  \frac{1}{\sqrt{2} \pi} \left( \frac{\kappa + u \alpha}{\gamma} \right)^{1/2}.
\end{equation}
Here, we have used the definitions of \(y_*\) and \(\lambda_0\) from \cref{ymax} and \cref{lambda_0}. The key takeaway from this result is that the frequency of betatron oscillations decreases with the electron’s energy gain, following a \(\gamma^{-1/2}\) scaling.

We now turn our attention to the frequency of laser field oscillations at the electron's location. We can say that a forward-moving electron experiences these oscillations at a frequency defined by
\begin{equation} \label{omega prime 0}
    \omega' \equiv d \xi/dt,
\end{equation}
where \(\xi\) is the laser phase at the electron's position. In the case of relativistic forward motion, this frequency is significantly lower than the laser frequency \(\omega_0\). Additionally, the betatron oscillations introduce substantial modulation to \(\omega'\) by altering the electron's longitudinal velocity. It is therefore useful to introduce the average frequency \(\langle \omega' \rangle\), defined as
\begin{equation}
    \langle \omega' \rangle = \frac{1}{2 \pi} \int_{- \pi}^{\pi} \omega' \, d \psi,
\end{equation}
which represents the average value of \(\omega'\) over one full betatron oscillation. 

To derive the expression for \(\omega'\), we begin by using Eqs.~(\ref{xi}) and (\ref{drdt}) to obtain the following approximate expression:
\begin{equation}
    \omega' \approx \frac{\omega_0}{\gamma} \left( \gamma - \frac{p_x}{m_e c} + \gamma \delta u \right),  \label{omega prime}
\end{equation}
where only the linear term in \(\delta u\) is retained. As shown in \cref{S_exp-4}, the combination of the first two terms inside the brackets undergoes significant modulation during betatron oscillations. Substituting this into \cref{omega prime} gives the following expression for \(\omega'\) in terms of the electron \(\gamma\)-factor and the betatron phase \(\psi\):
\begin{equation} \label{omega prime v3 new}
    \frac{\omega'}{\omega_0} = \frac{1}{\gamma} \left[ (S + \gamma \delta u) \cos^2 \psi + \gamma \delta u \right].
\end{equation}
\Cref{omega prime v3 new} explicitly shows that \(\omega'\) is modulated by betatron oscillations. From this expression, we find that the average frequency is given by:
\begin{equation} \label{omega prime av}
    \frac{\langle \omega' \rangle}{\omega_0} = \frac{1}{2\gamma} \left[ S + 3 \gamma \delta u \right].
\end{equation}    
An important conclusion from this expression is that \(\langle \omega' \rangle\) exhibits a markedly different dependence on \(\gamma\) compared to \(\omega_{\beta}\). Moreover, unlike \(\omega_{\beta}\), the behavior of \(\langle \omega' \rangle\) is non-monotonic.

To conclude this section, we express the laser phase $\xi$ in terms of the betatron phase $\psi$ using the derived expressions for $\omega_{\beta}$ and $\omega'$. Multiplying both sides of \cref{omega prime 0} by $dt$ and then substituting $dt = d \psi / \omega_{\beta}$, as derived from \cref{def: omega beta}, we find
\begin{equation}
    d \xi = \frac{\omega' (\psi)}{\omega_{\beta}} d \psi.
\end{equation}
Integrating over $\psi$, we obtain
\begin{equation}  \label{eq: 35}
    \xi(\psi) = \xi (\psi_0) + \int_{\psi_0}^{\psi} \frac{\omega'(\psi')}{\omega_{\beta}} d \psi'.
\end{equation}
The laser phase inherits the modulations of $\omega'$. In order to make this explicit, we re-write \cref{eq: 35} as
\begin{equation} \label{eq: xi v2}
    \xi(\psi) = \xi (\psi_0) + \frac{\langle \omega' \rangle}{\omega_{\beta}} (\psi - \psi_0) + \int_{\psi_0}^{\psi} \frac{\omega'(\psi') - \langle \omega' \rangle}{\omega_{\beta}} d \psi'.
\end{equation}
Here the second term on the right-hand side represents the modulations. The derived expression explicitly shows that for $\langle \omega' \rangle = l \omega_{\beta}$, the laser phase increases by $2 \pi l$ with each betatron oscillation, where $l$ is an integer. These resonances are the focus of this work, with $l = 1$ corresponding to the conventional betatron resonance.

%%%%%%%%%%%%%%%%%%%%%%%%%%%%%%%%%%%%%%%%%%%%%%%%%%%%%
%%%%%%%%%%%%%%%%%%%%%%%%%%%%%%%%%%%%%%%%%%%%%%%%%%%%%
%%%%%%%%%%%%%%%%%%%%%%%%%%%%%%%%%%%%%%%%%%%%%%%%%%%%%
%%%%%%%%%%%%%%%%%%%%%%%%%%%%%%%%%%%%%%%%%%%%%%%%%%%%%
%%%%%%%%%%%%%%%%%%%%%%%%%%%%%%%%%%%%%%%%%%%%%%%%%%%%%

\begin{figure}[h]
    \begin{center}
    \includegraphics[width=1\columnwidth,clip]{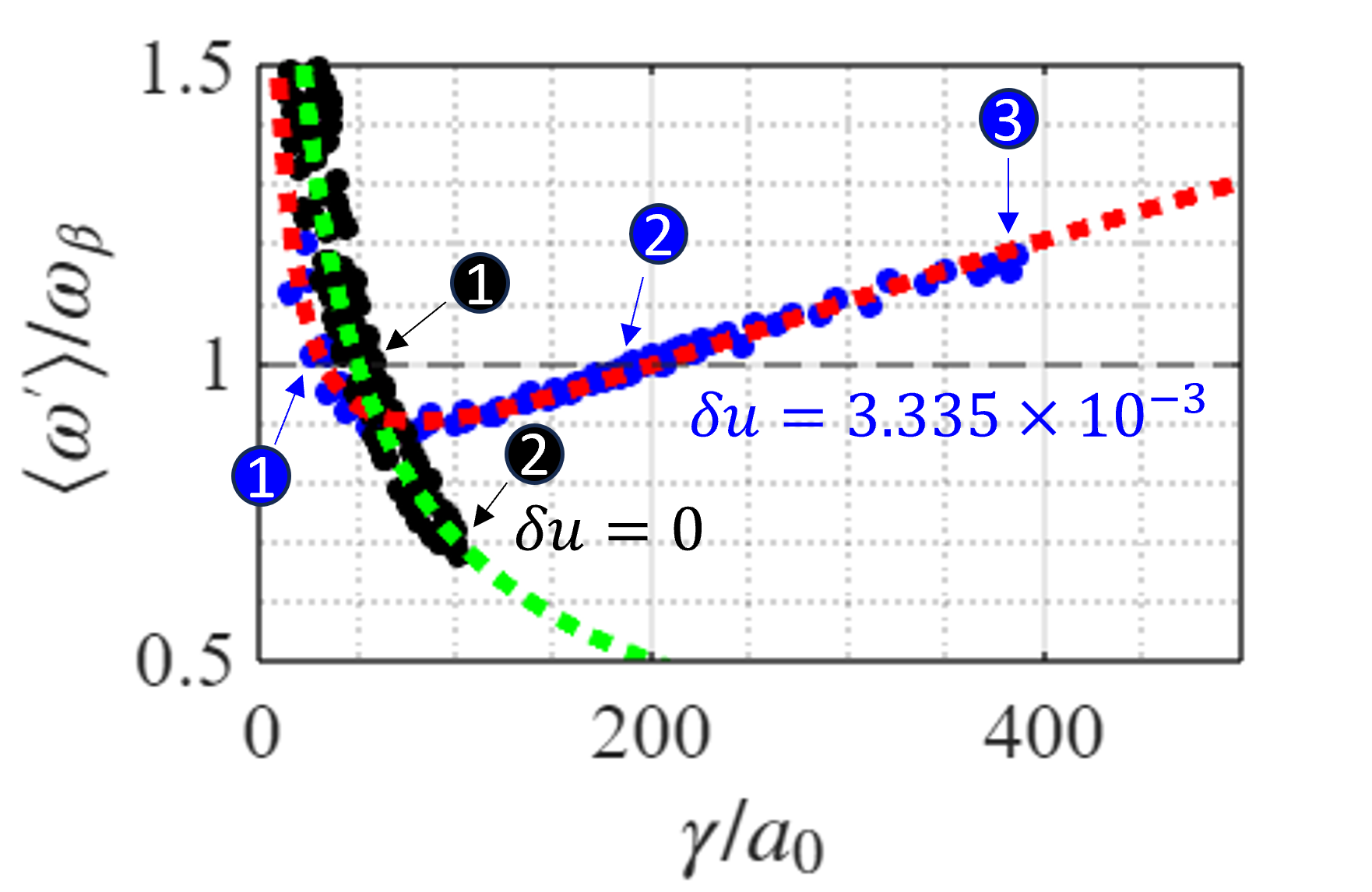}
    \caption{\label{fig:freq_ratio_1st} The frequency ratio \(\langle \omega' \rangle / \omega_{\beta}\) as a function of the relativistic \(\gamma\)-factor of the electron. The dashed curves represent the analytical expression given by \cref{omega prime over omega beta}, while the dots show values obtained from solving the equations of motion [Eqs.~(\ref{dpdt}) and (\ref{drdt})]. The green dashed curve and black dots correspond to \(\delta u = 0\) and \(S = 13.9394\), while the red dashed curve and blue dots correspond to \(\delta u = 3.335 \times 10^{-3}\) and \(S = 8.0303\). Numbered colored markers indicate key points discussed in the text: black (1), blue (1), and blue (2) correspond to the betatron resonance where \(\langle \omega' \rangle / \omega_{\beta} = 1\), while black (2) and blue (3) mark the points where the electron exits the energy-gaining phase. 
} 
    \end{center}
\end{figure}

\section{Frequency matching in betatron resonance} \label{sec: 1stres}

In \cref{sec: frequencies}, we derived analytical expressions for the two key frequencies, $\omega'$ and $\omega_{\beta}$, demonstrating that the laser phase increases by $2 \pi$ with each betatron oscillation under the condition of frequency matching, $\langle \omega' \rangle =  \omega_{\beta}$, referred to as the betatron resonance. Since both frequencies depend on electron energy, electron energy gain generally leads to frequency detuning. In this section, we show that this detuning can be mitigated by the superluminal phase velocity of the laser.

We begin this section by revisiting the expression for the net energy gain over one betatron oscillation, given by \cref{energy exchange 2}. Without loss of generality, we consider a betatron oscillation that starts at $\psi = -\pi$, corresponding to the electron being on the axis ($y = 0$). Using the expression for $v_y$ from \cref{v_y} and the expression for $\omega_{\beta}$ from \cref{def: omega beta}, we rewrite \cref{energy exchange 2} as 
\begin{equation} \label{eq:38}
    \Delta \gamma = - \frac{\Delta_{\max}}{\pi} \int_{-\pi}^{\pi} \cos (\psi) \cos (\xi) d\psi,
\end{equation}
where 
\begin{equation}
    \Delta_{\max} \equiv \pi a_0 y_* \omega_0/c.
\end{equation}

To illustrate the physics, consider the simple case where the laser phase has no modulations and the frequency condition for the betatron resonance is satisfied, $\langle \omega' \rangle = \omega' = \omega_{\beta}$. In this scenario, $\xi = \xi_0 + \psi $, where $\xi_0$ is the phase offset, so the expression for $\Delta \gamma$ given by \cref{eq:38} reduces to
\begin{equation} \label{eq:40}
    \Delta \gamma = - \frac{\Delta_{\max}}{\pi} \cos (\xi_0) \int_{-\pi}^{\pi} \cos^2 (\psi)  d\psi = -\Delta_{\max}  \cos (\xi_0).
\end{equation}
It is now clear that $\Delta_{\max}$ represents the maximum possible increase in $\gamma$ over one betatron oscillation, hence the subscript ``max''. We also see that the electron experiences a net energy gain only if the phase offset $\xi_0$ lies between $\pi/2$ and $3 \pi/2$. 

In actuality, modulations of $\omega'$ are always present. While they complicate the calculation (this can be found in Ref.~[\onlinecite{arefiev.pop.2024}]), they do not alter the key conclusion. The betatron resonance enables net energy gain during each betatron oscillation, provided that the phase offset is between $\pi/2$ and $3 \pi/2$. To be explicit, we define the phase offset as the difference between $\xi$ and $\psi$ evaluated at $\psi = 0$, corresponding to the electron being on the axis ($y=0$) in the middle of the considered betatron oscillation.

The net energy gain enabled by the frequency matching, in turn, affects the frequency matching itself, creating a feedback loop. It follows from 
\cref{eq: omega beta 0} and \cref{omega prime av} that the frequency ratio depends on $\gamma$:
\begin{equation} \label{omega prime over omega beta}
   \langle \omega' \rangle \left/\omega_{\beta} \right. \approx \chi \left( S \gamma^{-1/2} + 3 \gamma^{1/2} \delta u \right),
\end{equation}
where we define
\begin{equation} \label{eq:chi}
    \chi = \frac{\pi}{\sqrt{2 [\kappa + \alpha u]}}
\end{equation}
for compactness. In the case of luminal phase velocity ($\delta u = 0$), the frequency ratio decreases monotonically with increasing $\gamma$. This behavior is  illustrated in \cref{fig:freq_ratio_1st}  (dashed green curve) for an electron with $S = 13.9394$. The superluminal phase velocity ($\delta u > 0$) introduces a non-monotonic dependence of the frequency ratio on $\gamma$, driven by bifurcation-like sensitivity. This behavior is depicted in \cref{fig:freq_ratio_1st} (dashed red curve) for an electron with $S = 8.0303$, irradiated by a laser with $\delta u = 3.335 \times 10^{-3}$. The choice of these specific examples allows for a direct comparison between the cases that achieve the highest energy gain in the luminal and superluminal scenarios. These cases are identified based on the parameter scan that will be presented later in \cref{sec: scan}.

As evident from \cref{omega prime over omega beta}, electron energy gain can lead to frequency detuning. This detuning is detrimental, as it alters the phase offset between $\xi$ and $\psi$. Even small changes in the offset can accumulate over multiple betatron oscillations, eventually driving the phase offset into a range of values where the particle loses energy. In the case of luminal phase velocity, there is little that can be done to mitigate this behavior because the frequency ratio decreases monotonically with increasing $\gamma$. However, the non-monotonic dependence introduced by superluminal phase velocity provides an opportunity to reduce the detuning. This can be achieved by selecting the value of $S$ such that the betatron resonance condition is satisfied near the global minimum of $\langle \omega' \rangle / \omega_{\beta} $.

The frequency ratio and the corresponding value of $\gamma$ at the global minimum both depend on $\delta u$. We find from \cref{omega prime over omega beta} that the global minimum of the frequency ratio occurs at $\gamma = S / 3 \delta u$, with
\begin{equation} \label{frc}
     \min \left( \langle \omega' \rangle / \omega_{\beta} \right) = 2 \sqrt{3} \chi \left[ S \delta u \right]^{1/2}.
\end{equation}
This global minimum satisfies the betatron resonance condition, $\langle \omega' \rangle = \omega_{\beta}$, for
\begin{equation} \label{S^1}
    S = S_*^{(1)} \equiv \frac{1}{12 \chi^2} \delta u^{-1}.
\end{equation}
The corresponding value of $\gamma$ is thus
\begin{equation} \label{gamma_*^1}
    \gamma_*^{(1)} = \frac{1}{36 \chi^2} \delta u^{-2}.
\end{equation}
Since the value of $\min \left( \langle \omega' \rangle / \omega_{\beta} \right)$ given by \cref{frc} increases with $S$, electrons with $S > S_*^{(1)}$ never reach the betatron resonance. In such cases, $\langle \omega' \rangle / \omega_{\beta}$ remains greater than unity for all values of $\gamma$.

\begin{figure*}[!htb]
    \begin{center}
    \includegraphics[width=0.8\textwidth]{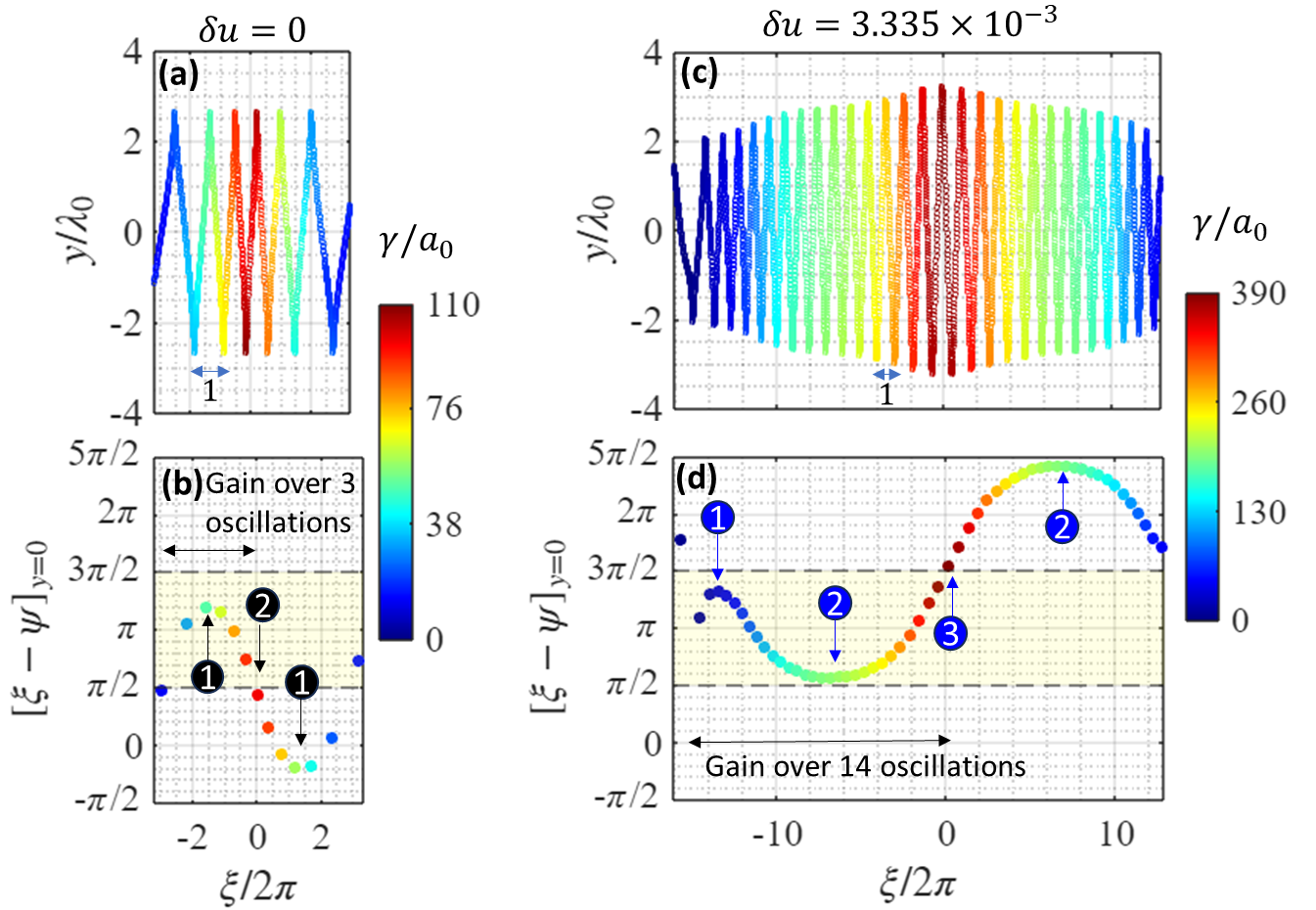}
    \caption{\label{fig:splm_lml} Electron transverse oscillations and phase offset for \(\delta u = 0\) (left column) and \(\delta u = 3.335 \times 10^{-3}\) (right column). Panels (a) and (c) show the electron’s transverse displacement as a function of the laser phase \(\xi\), with color representing the normalized \(\gamma\)-factor. Panels (b) and (d) display the evolution of the phase offset on the axis, \([\xi - \psi]_{y=0}\), with numbered markers indicating key points discussed in the text and shown in \cref{fig:freq_ratio_1st}. The yellow-shaded region highlights the phase offset range favorable for energy gain. For each case, the laser phase \(\xi\) has been offset by removing multiples of \(2\pi\) so that \(\xi = 0\) approximately corresponds to the point where the electron reaches its maximum energy. This adjustment, allowed since \(\xi\) appears in periodic functions, makes it easier to track the number of laser oscillations during the energy gain process.  }
    \end{center}
\end{figure*}

We now illustrate the impact of frequency detuning on electron energy gain for the two cases discussed earlier: one with a luminal phase velocity ($\delta u = 0$ and $S=13.9394$) and the other with a superluminal phase velocity ($\delta u = 3.335 \times 10^{-3}$ and $S = 8.0303$). For brevity, we will refer to these as the luminal and superluminal cases in this section. The electron dynamics in each case is determined by numerically solving the equations of motion given by Eqs.~(\ref{dpdt}) and (\ref{drdt}). In both cases, the electron starts at rest with a specific initial transverse displacement, $y_i$, which is chosen to set $S$ according to \cref{S_exp} as $S  = 1 +  [\kappa + u \alpha] y_i^2 / \lambda_0^2$. 

\Cref{fig:splm_lml}(a) and  \cref{fig:splm_lml}(c) show the numerically calculated transverse displacement as a function of the laser phase, \(\xi\), for the luminal and superluminal cases. The color represents the \(\gamma\)-factor normalized to \(a_0\). In both cases, there is a clear net energy gain with each betatron oscillation. In the superluminal case, the energy gain persists for approximately 14 oscillations, with the electron achieving \(\gamma / a_0 \approx 390\). In contrast, the energy gain in the luminal case lasts for only about 3 betatron oscillations, with the maximum \(\gamma\)-factor being roughly three times lower. This observation indicates that the electron in the superluminal case can maintain frequency matching over a much broader range of \(\gamma\), avoiding significant detuning. 

While we are interested in both the two key frequencies and the phase offset that they generate, these quantities are not directly provided by the equations of motion. Moreover, there is no explicit expression for computing the betatron phase, $\psi$, along the electron trajectory. However, by monitoring the electron's axis crossings, we can compute the frequency ratio and the phase offset twice during each betatron oscillation. This approach leverages the fact that each time the electron returns to the axis, the phase $\psi$ increases by $\pi$.

To compute the phase offset on the axis, $[\xi-\psi]_{y=0}$, we begin by selecting a reference point where the electron crosses the axis while moving upward. The laser phase at this location, denoted as $\xi_R$, is known, and we set $\psi = 0$ at this reference point. Therefore, the phase offset at this location is $[\xi-\psi]_{y=0} = \xi_R$. After half a betatron oscillation, the change in the phase offset is given by
\begin{equation}
    \Delta [\xi-\psi]_{y=0} = \Delta \xi_{\text{half}} - \pi,
\end{equation}
where $\Delta \xi_{\text{half}}$ is the corresponding change in $\xi$. By adding this value to the offset at the previous axis crossing, we compute the phase offset at subsequent points. Repeating this procedure allows us to determine the phase offset at all axis crossings along the electron trajectory.

To calculate the frequency ratio, \(\langle \omega' \rangle / \omega_{\beta}\), we again use the change in the wave phase at the axis crossings. \Cref{eq: xi v2} describes how the wave phase, \(\xi\), evolves with the betatron phase, $\psi$. At the end of one complete betatron oscillation, \(\psi\) increases by \(2\pi\), and the integral on the right-hand side of the equation vanishes. Consequently, the change in the wave phase over one full oscillation is given by \(\Delta \xi_{\text{full}} = 2\pi \langle \omega' \rangle / \omega_{\beta}\). By tracking the values of $\xi$ at the axis crossings, this change can be calculated numerically and then used to compute the frequency ratio as \(\langle \omega' \rangle / \omega_{\beta} = \Delta \xi_{\text{full}} / 2\pi\). 

As the number of oscillations during the energy gain process may not always be very large, the described method of computing the frequency ratio limits the number of data points that can be generated. To address this, we use the change in phase between two consecutive axis crossings, corresponding to half a betatron oscillation. During this interval, the change in \(\psi\) is \(\pi\), and the frequency ratio is given by  
\begin{equation}  
    \langle \omega' \rangle / \omega_{\beta} = \Delta \xi_{\text{half}} / \pi,  
\end{equation}  
where we assume that the integral on the right-hand side of \cref{eq: xi v2} vanishes when the integration is performed between two consecutive axis crossings. This assumption holds due to the symmetry of the betatron oscillations, as discussed in Ref.~[\onlinecite{arefiev.pop.2024}].  
%further supporting the observation of reduced detuning in the superluminal case

\Cref{fig:freq_ratio_1st} shows the frequency ratio calculated for both cases using the  described procedure. For the luminal case, the ratio monotonically decreases with \(\gamma\) (black dots), closely matching the analytical prediction shown as a green dashed curve. The energy gain ceases when the ratio drops below 0.7. In the superluminal case, the ratio is shown with blue dots and also matches the analytical prediction (red dashed curve). The plot confirms that the frequency detuning is indeed significantly slower, as the frequency ratio remains within 0.1 of \(\langle \omega' \rangle / \omega_{\beta} = 1\) until \(\gamma / a_0\) reaches 300.

\begin{figure}[h]
    \begin{center}
    \includegraphics[width=0.85\columnwidth,clip]{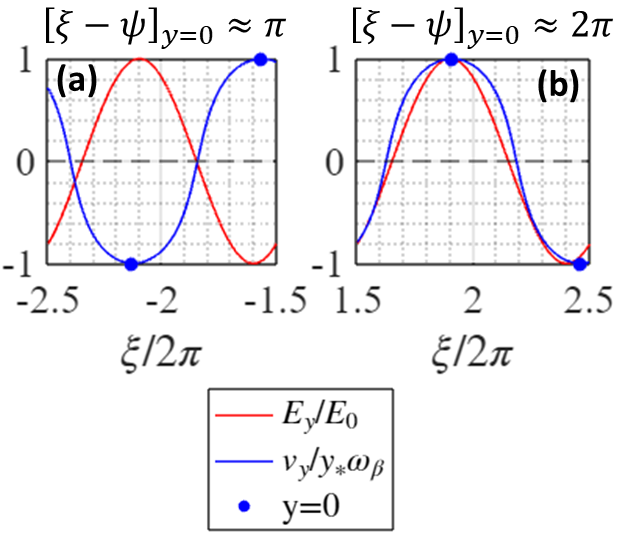}
    \caption{\label{fig:phase_diff} Evolution of the transverse laser electric field \( E_y \) and transverse electron velocity \( v_y \) during the superluminal \emph{betatron resonance} for two different values of the phase offset: (a) \([\xi - \psi]_{y=0} \approx \pi\) and (b) \([\xi - \psi]_{y=0} \approx 2\pi\). In panel (a), \( E_y \) is anti-parallel to \( v_y \) for most of the oscillation, leading to net energy gain. In panel (b), \( E_y \) is parallel to \( v_y \) for most of the oscillation, leading to net energy loss. The \( E_y \) and \( v_y \) curves are obtained from the same superluminal case shown in \cref{fig:splm_lml}(c) and \cref{fig:splm_lml}(d). The blue circles mark the points where the electron crosses the axis ($y=0$). } 
    \end{center}
\end{figure}

\Cref{fig:splm_lml}(b) and \cref{fig:splm_lml}(d) illustrate how the phase offset on the axis evolves with \(\xi\) for the luminal and superluminal cases. As shown earlier in this section, the phase offset range favorable for net energy gain lies between \(\pi/2\) and \(3\pi/2\), which is highlighted in yellow. In both cases, the electron gains energy within this region and loses energy outside it, as indicated by the color of the markers. \Cref{fig:phase_diff} further demonstrates the role of the phase offset in energy gain for the superluminal case by showing the evolution of \(E_y\) and \(v_y\) for two different values of \([\xi - \psi]_{y=0}\). When \([\xi - \psi]_{y=0} \approx \pi\) [\cref{fig:phase_diff}(a)], \(E_y\) is completely out of phase with \(v_y\), leading to energy gain. In contrast, when \([\xi - \psi]_{y=0} \approx 2\pi\) [\cref{fig:phase_diff}(b)], \(E_y\) and \(v_y\) are in phase, resulting in energy loss.  

To understand the difference in the evolution of the phase offset between the two cases, the phase offset must be considered in conjunction with the evolution of the frequency ratio, which is driven by changes in \(\gamma\). To track the frequency ratio, we mark the points where it reaches specific values using numbered circles — black for the luminal case and blue for the superluminal case. These markers are shown in \cref{fig:splm_lml}(b) and \cref{fig:splm_lml}(d), which track the evolution of the phase offset, as well as in \cref{fig:freq_ratio_1st}, which shows the dependence of the frequency ratio on \(\gamma\).  

We start by examining the luminal case, with the evolution of the phase offset plotted in \cref{fig:splm_lml}(b). The numbered black circles (1) indicate where \(\langle \omega' \rangle / \omega_{\beta} = 1\). To the left of the first marker (1), \(\langle \omega' \rangle / \omega_{\beta} > 1\), and the phase offset is within the range favorable for net energy gain. As a result, both the phase offset and \(\gamma\) increase with \(\xi\). This increase in \(\gamma\) drives the frequency ratio downward, as shown in \cref{fig:freq_ratio_1st}. Once \(\langle \omega' \rangle / \omega_{\beta}\) drops below unity, the phase offset starts to decrease. This corresponds to the region past the first marker (1) in \cref{fig:splm_lml}(b). At this stage, the phase offset remains within the favorable range for net energy gain, and \(\gamma\) continues to rise. The increase in \(\gamma\) persists until the phase offset drops below \(\pi/2\), at which point the electron exits the favorable range for net energy  gain, and energy gain ceases. The electron reaches its maximum \(\gamma\) when the phase offset reaches \(\pi/2\), marked by a numbered black circle (2) in \cref{fig:splm_lml}(b). This point corresponds to the rightmost point in \cref{fig:freq_ratio_1st}. Beyond marker (2) in \cref{fig:splm_lml}(b), the electron begins to lose energy, and \(\langle \omega' \rangle / \omega_{\beta}\) starts increasing again. Eventually, \(\langle \omega' \rangle / \omega_{\beta}\) returns to unity, and the phase offset starts increasing once more, as seen to the right of the second marker (1) in \cref{fig:splm_lml}(b).

We now perform a similar analysis for the superluminal case. The key difference compared to the luminal case is that \(\langle \omega' \rangle / \omega_{\beta} = 1\) occurs at two distinct values of \(\gamma\), as shown in \cref{fig:freq_ratio_1st}. The lower \(\gamma\) resonance is marked with a numbered blue circle (1), while the higher \(\gamma\) resonance is marked with a numbered blue circle (2). If the electron reaches the second (higher-$\gamma$) resonance marked with (2), the phase offset, instead of continuing to decrease as in the luminal case, begins to increase again, allowing further energy gain. As seen in \cref{fig:splm_lml}(d), the phase offset initially decreases after \(\langle \omega' \rangle / \omega_{\beta}\) drops below unity, similar to the luminal case. This behavior is observed to the right of the numbered blue circle (1). However, as \(\gamma\) continues to increase, \(\langle \omega' \rangle / \omega_{\beta}\) reaches its minimum and then rises, eventually returning to unity at the numbered blue circle (2). At this point, the phase offset stops decreasing and starts increasing, keeping the electron in the phase range favorable for energy absorption. As a result, the electron gains significantly more energy than in the luminal case and only exits this range when the phase offset exceeds \(3 \pi / 2\), marked with a numbered blue circle (3), which also corresponds to the rightmost point in \cref{fig:freq_ratio_1st}.  

\rc{To conclude this section, we revisit the key assumption $p_x \gg |p_y| \gg m_e c$, used in \cref{sec: frequencies} to derive the analytical expressions that underpin our interpretation of the results. This assumption, given by \cref{cond: 1}, is often referred to as the paraxial approximation. To evaluate its validity for the betatron resonance example shown in \cref{fig:splm_lml}(c), we plotted the ratio $|p_y|/p_x$ at instances when the electron crosses the axis ($y = 0$), as this is where the ratio reaches its highest value for each betatron oscillation. The plot is shown in \cref{appe: paraxial}, \cref{fig:paraxial}(a). The ratio $|p_y|/p_x$ decreases significantly over the course of the energy gain process, indicating that the condition $p_x \gg |p_y|$ is increasingly well satisfied. This confirms that the theory becomes applicable well before the electron reaches its peak energy and is well-suited for describing the electron dynamics.}

%%%%%%%%%%%%%%%%%%%%%%%%%%%%%%%%%%%%%%%%%%%%%%%%%%%%%
%%%%%%%%%%%%%%%%%%%%%%%%%%%%%%%%%%%%%%%%%%%%%%%%%%%%%
%%%%%%%%%%%%%%%%%%%%%%%%%%%%%%%%%%%%%%%%%%%%%%%%%%%%%
%%%%%%%%%%%%%%%%%%%%%%%%%%%%%%%%%%%%%%%%%%%%%%%%%%%%%
%%%%%%%%%%%%%%%%%%%%%%%%%%%%%%%%%%%%%%%%%%%%%%%%%%%%%

\section{Higher-order resonances} \label{sec: hres}

In the previous section, we examined the betatron resonance, which enables net energy gain over a single betatron oscillation. The key feature of this resonance is the frequency matching condition $\langle \omega' \rangle = \omega_{\beta}$. In this section, we explore higher-order resonances characterized by $\langle \omega' \rangle =  \ell \omega_{\beta}$, where $\ell$ is an odd integer. We show that these resonances can also benefit from a superluminal phase velocity, similarly to the betatron resonance.

Higher-order resonances differ fundamentally from the betatron resonance because they rely on modulations of $\omega'$ to gain energy. This can be seen explicitly by considering \cref{eq:38}, which gives the net energy gain over one betatron oscillation. In the absence of modulation, the laser phase is given by $\xi = \ell \psi + \xi_0$, where $\xi_0$ is the phase offset. Substituting this into \cref{eq:38}, we find that $\Delta \gamma = 0$ for $\ell > 1$, regardless of the phase offset. Thus, there is no net energy gain without modulation.   

An energy gain analysis that accounts for the frequency modulation induced by the betatron oscillation has already been carried out in Ref.~[\onlinecite{arefiev.pop.2024}]. Here, we simply present the result. It was shown that, once this modulation is included, the change in the relativistic factor 
$\gamma$ over one betatron oscillation no longer vanishes for odd $\ell$. For specified $\ell$ and phase offset $\xi_0$, the change is given by
\begin{equation} \label{delta-gamma-general}
    \frac{\left(\Delta \gamma\right)_{\ell}}{\Delta_{\max}} = (-1)^{k+1} \cos(\xi_0)  \left[ J_k (C_{\ell}) - J_{k+1} (C_{\ell}) \right],
\end{equation}
where 
\begin{eqnarray} \label{eq: C}
    && C_{\ell} =  \frac{\ell}{2} \frac{S + \gamma \delta u}{S + 3 \gamma \delta u}, \\
    && k = (\ell - 1)/2,
\end{eqnarray}
and $J_k$ and $J_{k+1}$ are the Bessel functions of the first kind.

We now examine this result for the specific case of $\ell = 3$. From Eq.~(\ref{eq: C}), we find that $0.5 < C_3 \leq 1.5$. In this range, $J_k (C_{\ell}) - J_{k+1} (C_{\ell}) = J_1 (C_3) - J_2 (C_3) > 0$, where we used the fact that $k = 1$ for $\ell = 3$. We therefore conclude that $\Delta \gamma > 0$ requires a phase offset in the range 
\begin{equation} \label{offset-3}
    -\pi/2 < \xi_0 < \pi/2.
\end{equation}

As in the case of betatron resonance, the net energy gain in a higher-order resonance also influences the frequency matching condition. Using \cref{eq: omega beta 0} and \cref{omega prime av}, we find that 
\begin{equation} \label{omega prime over omega beta - 3}
   \langle \omega' \rangle \left/ \ell \omega_{\beta} \right. \approx \chi \ell^{-1} \left( S \gamma^{-1/2} + 3 \gamma^{1/2} \delta u \right),
\end{equation}
which shows that the frequency ratio depends explicitly on $\gamma$, where $\chi$ is the quantity defined by \cref{eq:chi}. If the phase velocity is superluminal, the ratio exhibits a global minimum at $\gamma = S/3 \delta u$, with
\begin{equation} \label{frc-l}
     \min \left( \langle \omega' \rangle / \ell \omega_{\beta} \right) = 2 \sqrt{3} \chi \ell^{-1} \left[ S \delta u \right]^{1/2}.
\end{equation}

In the vicinity of the global minimum, the dependence of the frequency ratio on $\gamma$ is weakened, reducing the detuning from the resonance condition. The global minimum satisfies the frequency matching condition $\langle \omega' \rangle = \ell \omega_{\beta}$ for
\begin{equation} \label{S^l}
    S = S_*^{(\ell)} = \frac{1}{12 \chi^2} \frac{\ell^2}{\delta u}.
\end{equation}
As a result, the corresponding value of $\gamma$, given by $\gamma = S/3 \delta u$, scales as the square of the resonance order $\ell$:
\begin{equation} \label{gamma_*^l}
\gamma_*^{(\ell)} = \frac{\ell^2}{36 \chi^2} \delta u^{-2}.
\end{equation}
Equations (\ref{S^l}) and (\ref{gamma_*^l}) generalize the expressions derived for the betatron resonance, or $\ell = 1$, in \cref{sec: 1stres}.

\begin{figure}[h]
    \begin{center}
    \includegraphics[width=1\columnwidth,clip]{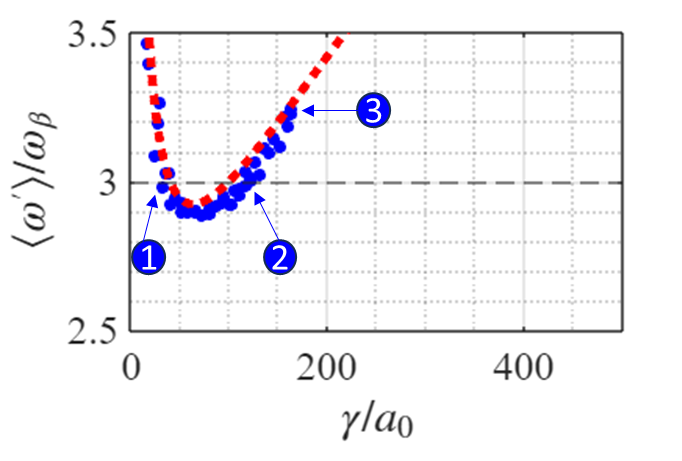}
    \caption{\label{fig:freq_ratio_3rd} The frequency ratio \(\langle \omega' \rangle / \omega_{\beta}\) as a function of the relativistic \(\gamma\)-factor for an electron with $S=23.15$ at $\delta u = 1.219 \times 10^{-2}$. The red dashed curve shows the analytical prediction given by \cref{omega prime over omega beta - 3} for $\ell = 3$, while the blue dots are obtained from numerical integration of the equations of motion [Eqs.~(\ref{dpdt}) and (\ref{drdt})]. Numbered blue markers indicate key points discussed in the text: (1) and (2) correspond to the third-order resonance where \(\langle \omega' \rangle / \omega_{\beta} = 3\); (3) marks the point where the electron exits the energy-gaining phase.} 
    \end{center}
\end{figure}

There are two key conclusions that follow from this result. The first conclusion is that electrons that cannot reach the betatron resonance may still access a higher-order resonance. As discussed in \cref{sec: 1stres}, electrons with $S > S^{(1)}_*$ cannot reach the betatron resonance, since the frequency ratio never decreases sufficiently. However, for fixed $\chi$ and $\delta u$, $S^{(\ell)}_* = \ell^2 S^{(1)}_*$, so $S^{(\ell)} > S^{(1)}_*$ for $\ell > 1$. As a result, electrons whose constant of motion $S$ is too large to access the betatron resonance may still experience a higher-order resonance of order $\ell$, provided that $S \leq S^{(\ell)}$.

The second conclusion is that higher-order resonances are likely to benefit from higher phase velocity. This follows from the behavior of the frequency ratio \( \langle \omega' \rangle / \ell \omega_\beta \), which has a minimum at \( \gamma = \gamma_*^{(\ell)} \), where the dependence on \( \gamma \) is weakest and frequency detuning is effectively mitigated. For fixed \( \chi \) and \( \delta u \), the value of \( \gamma_*^{(\ell)} \) increases with resonance order as \( \gamma_*^{(\ell)} = \ell^2 \gamma_*^{(1)} \), pushing this regime to higher energies for larger \( \ell \). As a result, accessing the detuning-mitigated regime becomes increasingly difficult. However, since \( \gamma_*^{(\ell)} \propto 1/\delta u^2 \), the required energy decreases as the phase velocity becomes more superluminal, making this regime more accessible even when \( \delta u \ll 1 \).

To conclude this section, we present an example that illustrates electron energy gain facilitated by the third-order resonance in a regime where the superluminal phase velocity reduces frequency detuning. We consider an electron with $S=23.15$ irradiated by a laser with $\delta u = 1.219 \times 10^{-2}$. The parameters are deliberately chosen to place the electron outside the range of the betatron resonance but within reach of the third-order resonance. Indeed, the electron's constant of motion $S$ is too large to allow access to the betatron resonance: $S > S_*^{(1)} \approx 2.72$, where $S_*^{(1)}$ is calculated using \cref{S^l}. However, since $S < S_*^{(3)} = 9 S_*^{(1)}\approx 24.44$, the electron may still experience the third-order resonance.

\Cref{fig:freq_ratio_3rd} shows the frequency ratio \(\langle \omega' \rangle / \omega_{\beta}\) as a function of the electron's relativistic factor $\gamma$. The red dashed curve represents the analytical prediction given by Eq.~\eqref{omega prime over omega beta - 3} for $\ell = 3$. We can see that the chosen parameters ensure that the detuning-mitigated regime is accessible for the third-order resonance. First, the frequency ratio at the local minimum is close to three, satisfying the resonance condition in a region where the dependence on $\gamma$ is weak. Second, the value of  $\gamma$ at the minimum is comparable to that in the betatron resonance example shown in \cref{fig:freq_ratio_1st}. This suggests that the regime remains accessible despite the higher resonance order. The key to achieving this is the use of a larger phase velocity (i.e., higher $\delta u$), which, according to the scaling in \cref{gamma_*^l}, lowers the value of $\gamma$ at the global minimum of the frequency ratio. This compensates for the increase in $\gamma$ associated with the $\ell^2$ scaling of higher-order resonances.

%The key to achieving this is the use of a larger phase velocity (i.e., higher $\delta u$), which reduces the required $\gamma$ according to the scaling given by \cref{gamma_*^l}, and offsets the increase associated with higher-order resonance. 

The blue dots in \cref{fig:freq_ratio_3rd} show values obtained from the numerical solution of the equations of motion [Eqs.~(\ref{dpdt}) and (\ref{drdt})]. The electron starts at rest with an initial transverse displacement, $y_i$, chosen to set $S$ according to \cref{S_exp} as $S  = 1 +  [\kappa + u \alpha] y_i^2 / \lambda_0^2$. We used the technique described in \cref{sec: 1stres} to compute \(\langle \omega' \rangle / \omega_{\beta}\). The numerical results are in good agreement with the analytical curve. We see that the electron is able to reach the third-order resonance [blue marker (1)] because this only requires $\gamma/a_0 \approx 40$. Once the electron reaches the resonance, it continues to gain energy while maintaining \(\langle \omega' \rangle / \omega_{\beta} \approx 3\). The energy gain eventually stops at $\gamma \approx 180 a_0$ [blue marker (3)]. The cutoff in energy gain is caused by the increase of \(\langle \omega' \rangle / \omega_{\beta} \) away from the resonant condition, as the electron moves beyond blue marker (2).

\begin{figure}[h]
    \begin{center}
    \includegraphics[width=1\columnwidth,clip]{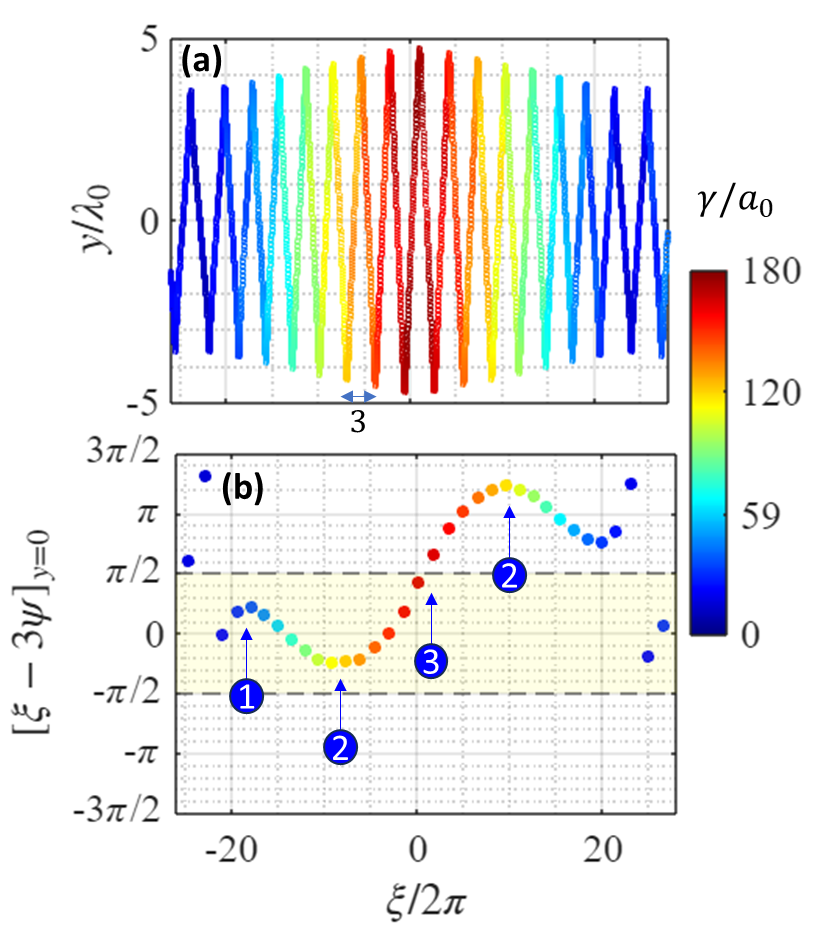}
    \caption{\label{fig:3rd_res} Electron transverse oscillations and phase offset for an electron with $S=23.15$ experiencing a third-order resonance in a laser with $\delta u = 1.219 \times 10^{-2}$. Panels (a) shows the electron’s transverse displacement as a function of the laser phase \(\xi\), with color representing the normalized \(\gamma\)-factor. Panel (b) displays the evolution of the phase offset on the axis, \([\xi - 3\psi]_{y=0}\), with numbered markers indicating key points discussed in the text and shown in \cref{fig:freq_ratio_3rd}. The yellow-shaded region highlights the phase offset range favorable for energy gain. The laser phase \(\xi\) has been offset by removing multiples of \(2\pi\) so that \(\xi = 0\) approximately corresponds to the point where the electron reaches its maximum energy. This adjustment, allowed since \(\xi\) appears in periodic functions, makes it easier to track the number of laser oscillations during the energy gain process. 
} 
    \end{center}
\end{figure}

\Cref{fig:3rd_res}(a) shows the numerically calculated transverse displacement of the same electron as a function of the laser phase \( \xi \). The color scale indicates the relativistic factor \( \gamma \) normalized to \( a_0 \). The plot shows clear net energy gain over successive betatron oscillations. During each oscillation, the electron experiences approximately three laser cycles, consistent with operation in the third-order resonance regime. The energy gain continues for about seven betatron oscillations — or roughly 21 laser cycles — allowing the electron to reach \( \gamma / a_0 \approx 180 \). The figure confirms that, in this regime, the electron maintains frequency matching over a broad range of \( \gamma \), thereby avoiding significant detuning.

To emphasize the crucial role of the minimum in the frequency ratio shown in \cref{fig:freq_ratio_3rd} for sustaining energy gain, we again examine the phase offset between the betatron oscillations and the laser field. In this case, we define the offset as \( [\xi - 3\psi]_{y=0} \), so that it remains constant when the electron is in third-order resonance. It can be shown, using the expressions given in Ref.~[\onlinecite{arefiev.pop.2024}], that when the modulation of the laser frequency by the betatron motion is taken into account, this offset indeed remains constant during third-order resonance. The offset is computed using a procedure similar to that described in \cref{sec: 1stres}.

\Cref{fig:3rd_res}(b) illustrates how the phase offset on the axis evolves with \( \xi \). As derived earlier [see \cref{offset-3}], the phase offset favorable for net energy gain during third-order resonance lies between \( -\pi/2 \) and \( \pi/2 \), highlighted by the yellow region. The electron indeed gains energy within this range and loses energy outside it, as indicated by the color of the markers. The key feature is that the phase offset reverses direction twice to remain within the favorable range: it initially increases, then decreases, and finally increases again. This behavior is similar to that shown in \cref{fig:splm_lml}(d) for the betatron resonance in the presence of a superluminal phase velocity.

To explain this behavior, we again consider the phase offset in conjunction with the frequency ratio, which evolves due to changes in \( \gamma \). To aid interpretation, we mark several key points — corresponding to specific values of the frequency ratio — using numbered blue circles. These markers appear both in \cref{fig:3rd_res}(b), which shows the evolution of the phase offset, and in \cref{fig:freq_ratio_3rd}, which shows the corresponding values of \( \langle \omega' \rangle / \omega_\beta \) as a function of \( \gamma \).

The key feature of \( \langle \omega' \rangle / \omega_\beta \) is that it matches the third-order resonance condition at two distinct values of \( \gamma \), as shown in \cref{fig:freq_ratio_3rd}. The lower-\( \gamma \) resonance is marked with marker (1), while the higher-\( \gamma \) resonance is marked with marker (2). To the left of marker (1), \( \langle \omega' \rangle / \omega_{\beta} > 3 \) and the phase offset is within the range favorable for energy gain. As a result, both the phase offset and \( \gamma \) increase with \( \xi \). This increase in \( \gamma \) drives the frequency ratio downward, as shown in \cref{fig:freq_ratio_3rd}. Once \( \langle \omega' \rangle / \omega_{\beta} \) drops below 3, the phase offset begins to decrease. This transition corresponds to the region past marker (1) in \cref{fig:3rd_res}(b).

Without the higher-\( \gamma \) resonance, the phase offset would continue to decrease and the net energy gain would eventually stop. However, as \( \gamma \) increases, \( \langle \omega' \rangle / \omega_{\beta} \) reaches its minimum and then begins to rise again, returning to the third-order resonance condition at marker (2). At this point, the phase offset stops decreasing and starts increasing, allowing the electron to stay within the phase range favorable for net energy gain. The electron finally exits this range when the phase offset exceeds \( \pi/2 \), marked by blue circle (3), which also corresponds to the rightmost point in \cref{fig:freq_ratio_3rd}. This sequence, with two reversals of the phase offset, allows the electron to maintain net energy gain over a broad range of \( \gamma \), similarly to the behavior observed for the betatron resonance in the presence of a superluminal phase velocity.

\rc{Similar to what we did for the betatron resonance in \cref{sec: 1stres}, we now revisit the key assumption $p_x \gg |p_y| \gg m_e c$, used in deriving the analytical expressions of \cref{sec: frequencies} that inform our interpretation of the results. To assess its validity for the third-order resonance example shown in \cref{fig:3rd_res}(a), we plotted the ratio $|p_y|/p_x$ at instances when the electron crosses the axis ($y = 0$), as this is where the ratio reaches its highest value for each betatron oscillation. The plot is shown in \cref{appe: paraxial}, \cref{fig:paraxial}(b). As in the betatron resonance case, the ratio $|p_y|/p_x$ decreases significantly during the energy gain process, indicating that the condition $p_x \gg |p_y|$ becomes increasingly well satisfied. This again confirms that the theory becomes applicable well before the electron reaches its peak energy and is well-suited for describing the electron dynamics.}

When comparing the results presented in this section to those in \cref{sec: 1stres}, we conclude that the third-order resonance, in the regime where the superluminal phase velocity reduces the frequency detuning by weakening the dependence of the frequency ratio on $\gamma$, offers a viable path to enhanced energy gain. Although the observed energy gain is lower than that achieved in the superluminal case with a betatron resonance shown in \cref{fig:freq_ratio_1st}, it significantly exceeds the energy gain obtained in the luminal case, also shown in \cref{fig:freq_ratio_1st}.

%%%%%%%%%%%%%%%%%%%%%%%%%%%%%%%%%%%%%%%%%%%%%%%%%%%%%
%%%%%%%%%%%%%%%%%%%%%%%%%%%%%%%%%%%%%%%%%%%%%%%%%%%%%
%%%%%%%%%%%%%%%%%%%%%%%%%%%%%%%%%%%%%%%%%%%%%%%%%%%%%
%%%%%%%%%%%%%%%%%%%%%%%%%%%%%%%%%%%%%%%%%%%%%%%%%%%%%
%%%%%%%%%%%%%%%%%%%%%%%%%%%%%%%%%%%%%%%%%%%%%%%%%%%%%

\section{Phase velocity scan} \label{sec: scan}

In \cref{sec: 1stres} and \cref{sec: hres}, we examined three representative examples that demonstrated the benefit of a superluminal phase velocity and its counterintuitive ability to reduce frequency detuning by weakening the dependence of the frequency ratio on \( \gamma \). In this section, we extend that analysis by performing a parameter scan focused on varying the degree of superluminosity \( \delta u \), while also sampling a range of values for the integral of motion \( S \). 

The scan is carried out by numerically solving the equations of motion given by Eqs.~(\ref{dpdt}) and~(\ref{drdt}). All parameters are held fixed except for \( \delta u \) and \( S \). As stated at the end of \cref{sec: test-model}, we set \( a_0 = 10 \), \( \alpha = 1.52 \), and \( \kappa = 0.44 \) for all numerical calculations. The initial phase is again set to \( \xi_* = \pi/2 \). In each case, the electron starts at rest with an initial transverse displacement \( y_i \), chosen to set \( S \) according to \cref{S_exp} as \( S = 1 + [\kappa + u \alpha] y_i^2 / \lambda_0^2 \). 

To ensure that the electron reaches its maximum energy for each parameter set, we run each simulation until the electron travels a fixed distance of \( 2000 \lambda_0 \) along the \( x \)-axis in all simulations. This distance is based on the case shown in \cref{fig:splm_lml}(c), where the electron attains its peak energy just before reaching this point. The actual distance required in that case is only slightly shorter. We verified, using several representative parameter sets, that increasing the travel distance to \( 5000 \lambda_0 \) and \( 20000 \lambda_0 \) yields virtually identical results, confirming that \( 2000 \lambda_0 \) is sufficient for capturing the maximum energy gain.

%To ensure a consistent comparison, we limit the longitudinal distance traveled by the electron. In each calculation, the electron is allowed to travel a distance of \( 2000 \lambda_0 \) along the \( x \)-axis.

\begin{figure}[h]
    \begin{center}
    \includegraphics[width=1\columnwidth,clip]{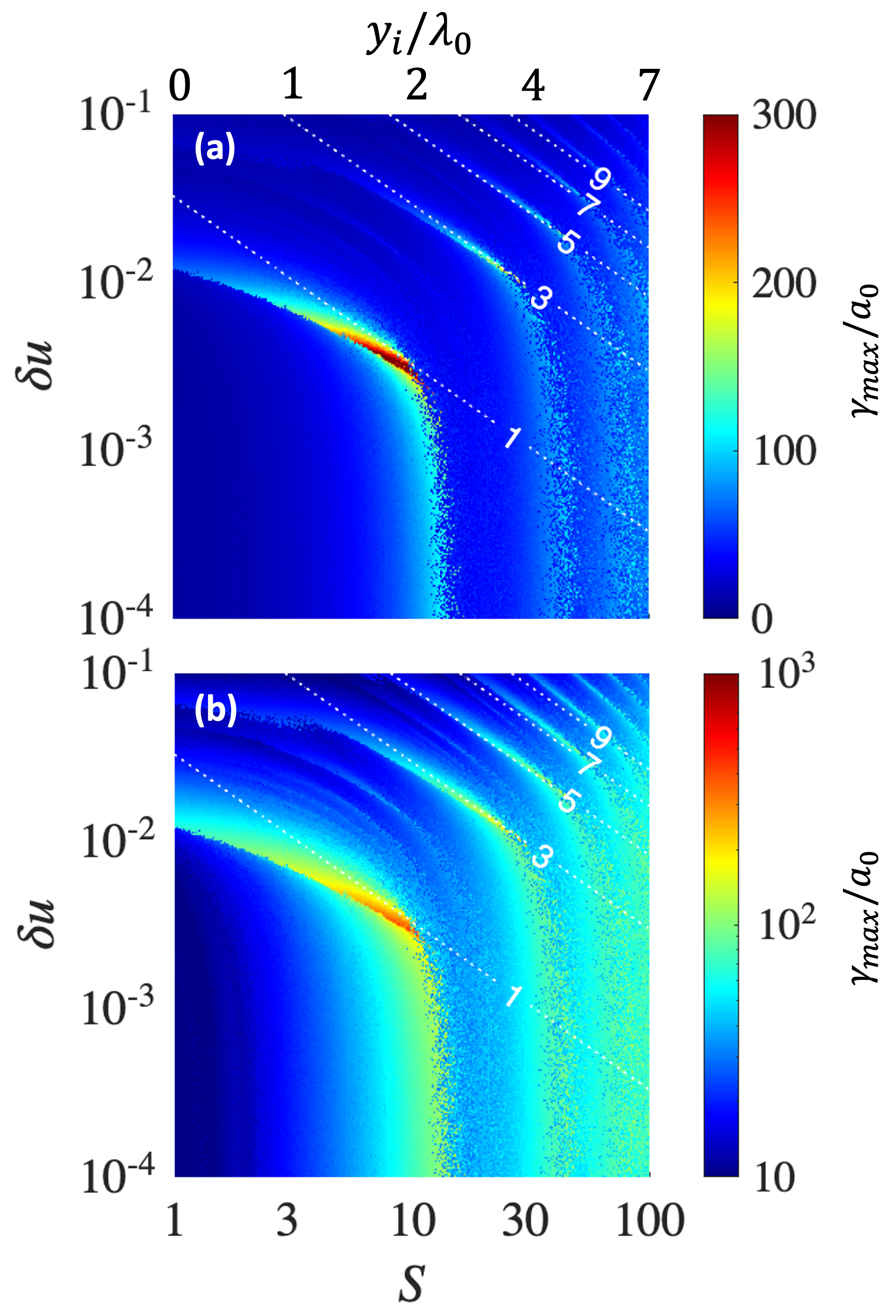}
    \caption{\label{fig:pl_2Dscan} The maximum relativistic factor, \( \gamma_{\max} \), attained by an electron while traveling a distance of \( 2000 \lambda_0 \) in the longitudinal direction. \rc{Panel (a) uses a linear color scale, while panel (b) uses a logarithmic scale to improve visibility of lower-energy regions.} Each pixel represents a numerical solution of Eqs.~(\ref{dpdt}) and~(\ref{drdt}) for an electron with a given value of \( S \) and a laser with a given relative degree of superluminosity \( \delta u \). All other laser and plasma parameters are held fixed throughout the scan (see main text for details). The numbers along the top of the plot indicate the initial transverse displacement \( y_i \), computed at \( \delta u = 0.01 \) for the corresponding value of \( S \) shown along the bottom axis. The dashed lines represent \( S_*^{(\ell)} \), given by Eq.~(\ref{S^l}), for odd \( \ell \), with each curve labeled by its corresponding \( \ell \) value.} 
    \end{center}
\end{figure}

\Cref{fig:pl_2Dscan} shows the result of a parameter scan covering the range $10^{-4} \leq \delta u \leq 10^{-1}$ with $1 \leq S \leq 100$. The color indicates the maximum relativistic factor, denoted $\gamma_{\max}$, attained by the electron. \rc{Panel (a) uses a linear color scale, while panel (b) uses a logarithmic scale to improve visibility of lower-energy regions.} The figure reveals distinct bands of enhanced energy gain distributed across the parameter space.

To aid in identifying each band, we overlay dashed curves in \cref{fig:pl_2Dscan} that represent \( S_*^{(\ell)} \), given by \cref{S^l}. These curves show, for each resonance order \( \ell \), the values of the integral of motion \( S \) for which the global minimum of the frequency ratio — considered as a function of \( \gamma \) — satisfies the frequency matching condition \( \langle \omega' \rangle / \omega_{\beta} = \ell \). The curves reflect the fact that the position of this minimum depends on the phase velocity through \( \delta u \), so the required value of $S$ varies with \( \delta u \) for fixed \( \ell \). Because efficient resonances are expected only for odd \( \ell \), we plot the curves for odd values only, labeling each with the corresponding resonance order. Although these curves are shown over the entire parameter range, their derivation assumes $\delta u \ll 1$, so they are only applicable in this range.

We now examine how the dashed curves relate to the bands of enhanced energy gain observed in \cref{fig:pl_2Dscan}. The \( \ell = 1 \) curve grazes one of the bands, with a region of peak energy gain located near the grazing point. The superluminal example from \cref{sec: 1stres}, which relies on the first-order (betatron) resonance, lies within this region. A similar correspondence is observed between the other dashed curves and the centers of nearby bands. In particular, the example from \cref{sec: hres}, which involves the third-order resonance, belongs to the region of peak energy gain where the \( \ell = 3 \) curve grazes the next band. Guided by these observations, we will refer to each band using the \( \ell \) value of the corresponding dashed curve.

While we have so far focused on relating each band to a corresponding curve, it is also important to emphasize that the bands do not follow the curves across the full parameter range. This deviation is particularly pronounced at lower values of \( \delta u \), which we refer to as the low-\( \delta u \) regime. At the opposite end, in what we refer to as the high-\( \delta u \) regime, the energy gain decreases with increasing \( \delta u \). For example, the \( \ell = 1 \) band noticeably deviates from the \( \ell = 1 \) curve in this regime, while the deviation for the \( \ell = 3 \) band is present but less pronounced. In what follows, we examine each regime separately.

\subsection{Low-$\delta u$ regime}

A key feature of the low-\(\delta u\) regime is that the location of each band becomes insensitive to further decreases in \(\delta u\). At the same time, the energy gain also becomes nearly independent of \(\delta u\). This behavior can be understood by examining the analytical expression for the frequency ratio \(\langle \omega' \rangle / \omega_{\beta}\) given in \cref{omega prime over omega beta - 3}. It follows from this expression that, for a fixed range of \(\gamma\), the dependence on \(\delta u\) disappears provided that
\begin{equation} \label{low delta u}
    \delta u \ll S / 3 \gamma_{\max}.
\end{equation}
In this limit, the frequency ratio simplifies to
\begin{equation} 
   \langle \omega' \rangle \left/ \omega_{\beta} \right. \approx \chi S \gamma^{-1/2}.
\end{equation}
This relation implies that, for a given \(S\), the frequency ratio becomes independent of \(\delta u\) over the relevant \(\gamma\) range, which in turn explains why the energy gain, which is controlled by the frequency ratio, saturates at low \(\delta u\).

The key conclusion from our analysis is that the phase velocity of the laser can effectively be treated as luminal in the low-\(\delta u\) regime. The onset of this regime depends on the band: higher-\(\ell\) bands correspond to larger values of \(S\), which in turn raise the threshold for entering the low-\(\delta u\) regime. For example, \(10 \lesssim S \lesssim 15\) in the \(\ell = 1\) band, where \(\gamma_{\max}/a_0 \lesssim 120\), whereas \(35 \lesssim S \lesssim 50\) in the \(\ell = 3\) band, where \(\gamma_{\max}/a_0 \lesssim 60\). It then follows from \cref{low delta u} that the low-\(\delta u\) regime extends to higher values of \(\delta u\) in the \(\ell = 3\) band, which is consistent with what is seen in \cref{fig:pl_2Dscan}.

\begin{figure}[h]
    \begin{center}
    \includegraphics[width=1\columnwidth,clip]{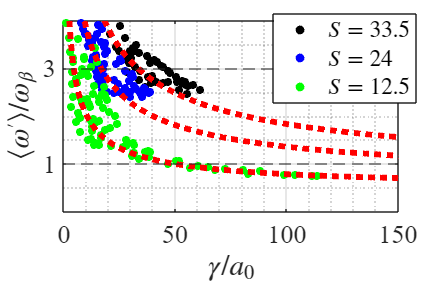}
    \caption{\label{fig:ratio_du_1e-3}  Frequency ratio \(\langle \omega' \rangle / \omega_{\beta}\) as a function of the relativistic factor \(\gamma\) in the low-$\delta u$ regime for three values of the integral of motion: $S = 12.5$, $S = 24$, and $S=33.5$, with $\delta u=10^{-3}$ in all cases. The dashed curves show the analytical prediction from \cref{omega prime over omega beta - 3}, evaluated for the three different values of $S$ at $\ell = 1$. The filled markers are obtained from numerical integration of the equations of motion [Eqs.~(\ref{dpdt}) and (\ref{drdt})]. Green, blue, and black markers correspond to $S = 12.5$, $S = 24$, and $S=33.5$.
} 
    \end{center}
\end{figure}

The scan in \cref{fig:pl_2Dscan} also shows that the values of \( S \) associated with a given \(\ell\) band lie well below the corresponding \( S_*^{(\ell)} \) predicted by the dashed lines. To understand this trend, we examine the frequency ratio for three representative values of \( S \) at a fixed phase velocity offset of \( \delta u = 10^{-3} \), as shown in \cref{fig:ratio_du_1e-3}. The first value, \( S = 12.5 \), corresponds to the location of the \(\ell = 1\) band in the scan. The third value, \( S = 33.5 \), corresponds to \( S_*^{(1)} \) at this value of \( \delta u \). The second value, \( S = 24 \), is chosen as an intermediate point between the two. \Cref{fig:ratio_du_1e-3} shows both the analytical prediction from \cref{omega prime over omega beta - 3} for \( \ell = 1 \) and the results from the numerical integration of the equations of motion [Eqs.~(\ref{dpdt}) and~(\ref{drdt})] for each value of \( S \). 

The analytical prediction highlights a key distinction between the \( S = 12.5 \) case and the other two. For \( S = 12.5 \), the frequency ratio drops below unity within the considered range of \( \gamma \), whereas for the other two values it remains above unity throughout. This drop below unity enables the phase offset to reverse its direction of change, keeping it within the range favorable for energy gain for longer.

A direct comparison of the numerical results for \( S = 12.5 \) (green markers) and \( S = 24 \) (blue markers) highlights the benefit of the frequency ratio dropping below unity. In the case of \( S = 24 \), the frequency ratio drops only to about 2.75 before energy gain terminates at around \( \gamma/a_0 \approx 30 \). In contrast, for \( S = 12.5 \), the frequency ratio approaches the resonance condition \( \langle \omega' \rangle / \omega_{\beta} = 1 \) at the same value of \( \gamma \). At this stage, the electron begins to benefit from the betatron resonance and continues gaining energy. As discussed earlier, the drop below unity allows the phase offset to reverse, which helps keep the electron in a favorable phase relationship with the wave for an extended time.

The case of \( S = 33.5 \) (black markers) is particularly notable because, although this value corresponds to the global minimum of the frequency ratio aligning with the betatron resonance condition, the electron instead benefits from the third-order resonance. As in the \( S = 24 \) case, the frequency ratio remains well above unity throughout the considered range of \( \gamma \). However, around \( \gamma/a_0 \approx 30 \), the ratio reaches the third-order resonance condition, \( \langle \omega' \rangle / \omega_{\beta} = 3 \). At this point, the electron begins to benefit from the third-order resonance and continues gaining energy. As \( \gamma \) increases, the frequency ratio drops below three, causing the phase offset defined for the third-order resonance to decrease. This evolution helps maintain a favorable phase relationship with the wave, extending the energy gain period. In this case, the electron reaches \( \gamma/a_0 \approx 60 \). Notably, this value of \( S \) lies within the \( \ell = 3 \) band in the scan shown in \cref{fig:pl_2Dscan}.

\subsection{High-$\delta u$ regime}

A key feature of the high-\(\delta u\) regime is the reduction in energy gain as \(\delta u\) increases. To understand this trend, we revisit Eq.~(\ref{gamma_*^l}), which gives the location, $\gamma_*^{(\ell)}$, of the global minimum of the frequency ratio \(\langle \omega' \rangle / \omega_{\beta}\) for a given resonance order \(\ell\). The equation shows that 
\[
\gamma_*^{(\ell)} \propto \ell^2 / \delta u^2,
\]
which means that the minimum shifts to lower values of \(\gamma\) as \(\delta u\) increases.

\begin{figure}[t]
    \begin{center}
    \includegraphics[width=1\columnwidth,clip]{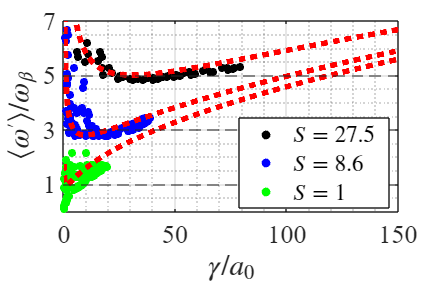}
    \caption{\label{fig:res_highdu} Frequency ratio \(\langle \omega' \rangle / \omega_{\beta}\) as a function of the relativistic factor \(\gamma\) at $\delta u=3 \times 10^{-2}$ for three values of the integral of motion: $S = 1$, $S = 8.6$, and $S=27.5$. The dashed curves show the analytical prediction from \cref{omega prime over omega beta - 3}, evaluated for the three different values of $S$ at $\ell = 1$, $\ell = 3$, and $\ell = 5$. The filled markers are obtained from numerical integration of the equations of motion [Eqs.~(\ref{dpdt}) and (\ref{drdt})]. Green, blue, and black markers correspond to $S = 1$, $S = 8.6$, and $S=27.5$.
} 
    \end{center}
\end{figure}

In the low-\(\delta u\) regime, this minimum is located at very high \(\gamma\), beyond what the electron can reach — rendering the minimum effectively irrelevant. As \(\delta u\) increases, the minimum moves into the accessible range of \(\gamma\), which initially enhances the energy gain. This trend reaches its optimum in the region of peak energy gain within each \(\ell\) band.

However, further increase in \(\delta u\) becomes counterproductive. As the minimum shifts to even lower values of \(\gamma\), the electron reaches it too early in its trajectory. Beyond this point, the frequency ratio begins to rise again with \(\gamma\), quickly taking the electron out of resonance. This early departure from the resonance condition limits the energy gain and explains the observed decrease in performance at high \(\delta u\).

To further illustrate the argument presented above, we plot the frequency ratio at a fixed \( \delta u = 3 \times 10^{-2} \) for three different values of \( S \). These values are chosen so that the global minimum of the frequency ratio \( \langle \omega' \rangle / \omega_{\beta} \), considered as a function of \( \gamma \), satisfies the resonance condition for \( \ell = 1 \), \( \ell = 3 \), and \( \ell = 5 \), respectively. This is evident from the analytical curves shown as dashed lines in \cref{fig:res_highdu}, each of which reaches its minimum at the corresponding resonance value. In the case of \( S = 1 \), the analytical curve reaches the betatron resonance condition rapidly, but then increases steeply. As a result, the calculated energy gain (green markers) remains relatively limited, with \( \gamma/a_0 \lesssim 20 \).

In the other two cases with \( S = 8.6 \) and \( S = 27.5 \), the local minima match the resonance conditions for the third- and fifth-order resonances, respectively. Since \( \gamma_*^{(\ell)} \propto \ell^2 \), higher-order resonances are better suited for enhancing electron energy gain in the high-\( \delta u \) regime. For \( S = 8.6 \), the electron benefits from the third-order resonance and is able to reach \( \gamma/a_0 \approx 40 \), which is twice the value achieved in the \( S = 1 \) case. However, the best performance is seen in the \( S = 27.5 \) case. The global minimum occurs at a value of \( \gamma \) that is still accessible, and the analytical curve remains relatively flat beyond the resonance compared to the other two cases. As a result, the electron benefits from the fifth-order resonance and reaches \( \gamma/a_0 \approx 80 \). Both the \( S = 1 \) and \( S = 8.6 \) cases exemplify the limitations of the high-\( \delta u \) regime, where premature detuning limits energy gain. In contrast, the \( S = 27.5 \) case avoids this limitation by benefiting from a broader and more gradual transition away from the resonance.

A key takeaway from this result is that, due to the \(\ell^2 / \delta u^2\) dependence of the location of the global minimum of the frequency ratio, the onset of the high-\(\delta u\) regime is delayed for higher-order resonances. As a consequence, the regions of peak energy gain for different \(\ell\) bands in \cref{fig:pl_2Dscan} become separated in \(\delta u\). We can thus conclude that higher-order resonances provide a viable mechanism for enhanced electron energy gain in regimes where the betatron resonance (\(\ell = 1\)) becomes ineffective due to increased values of \(\delta u\).

%**********************************

\section{Summary and discussion} \label{sec: summary}

This work examined electron dynamics and energy gain during direct laser acceleration in a regime where the laser’s phase velocity is superluminal and static plasma fields induce betatron oscillations. The central focus was on how frequency matching between the oscillating laser field and the betatron motion evolves with electron energy, and how this matching governs the net energy gain.

We derived analytical expressions for the betatron frequency and the average frequency of laser field oscillations experienced by the electron, and showed that a superluminal phase velocity introduces a non-monotonic dependence of their ratio on the relativistic factor \( \gamma \). This leads to a global minimum that can stabilize frequency matching over a broad energy range, allowing the electron to remain in resonance even as it gains energy. Notably, the location of this minimum depends on both the resonance order \( \ell \) and the relative degree of superluminosity \( \delta u \).

Using numerical solutions of the test-electron model, we demonstrated that the suppression of frequency detuning by a superluminal phase velocity enhances energy gain not only for the conventional betatron resonance (\( \ell = 1 \)) but also for higher-order resonances (\( \ell = 3, 5, \dots \)). We also performed a parameter scan by varying \( \delta u \) and the integral of motion \( S \). The scan revealed distinct bands of enhanced energy gain, each associated with a specific resonance order. Importantly, the regions of peak energy gain within these bands appear at different values of \( \delta u \). As \( \delta u \) increases, the betatron resonance becomes ineffective due to premature frequency detuning. At the same time, higher-order resonances become increasingly effective, emerging as the dominant mechanisms for enhanced energy gain in this regime of direct laser acceleration.

\begin{figure}[t]
    \begin{center}
    \includegraphics[width=1\columnwidth,clip]{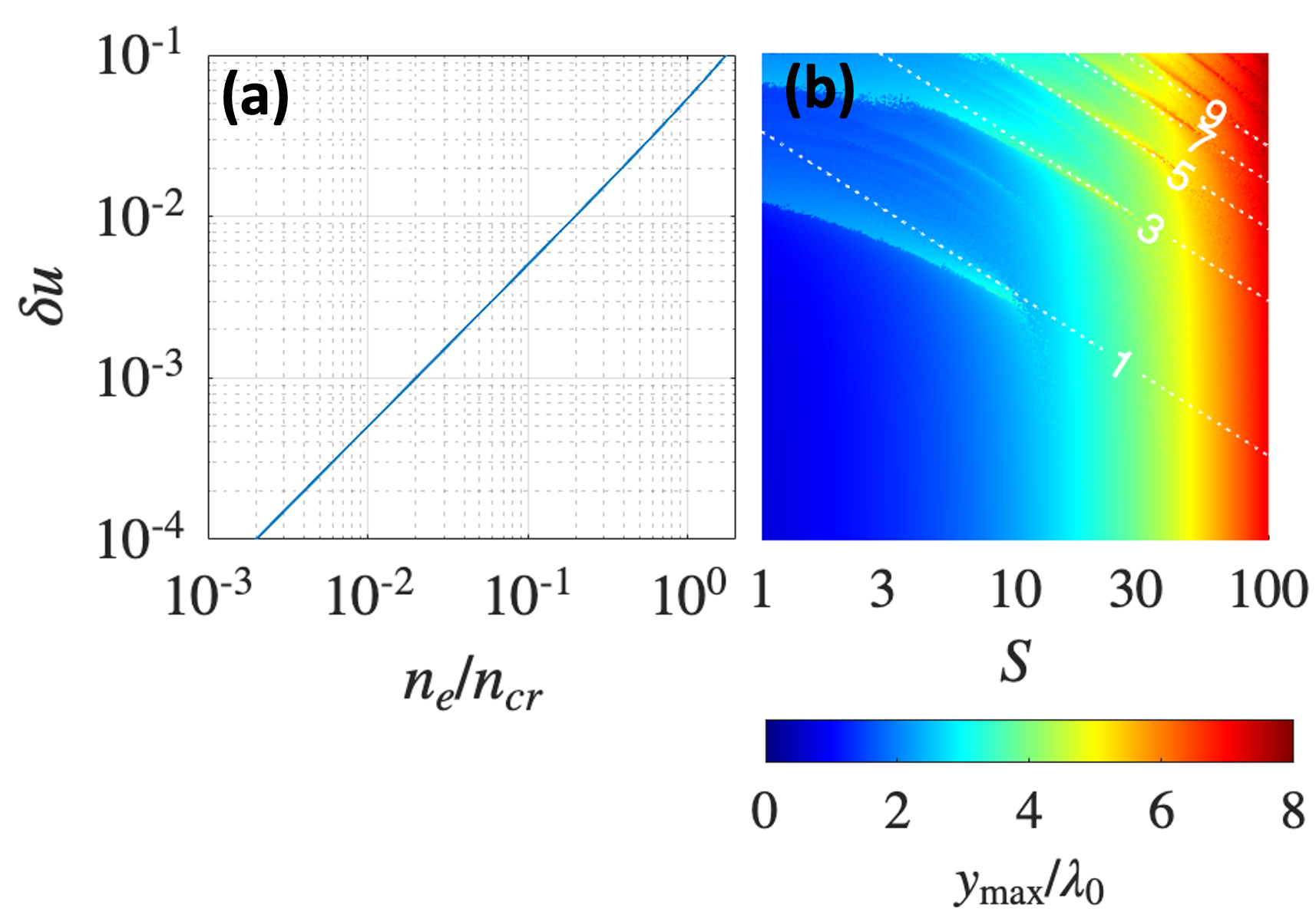}
    \caption{\label{fig:vph_ymax} \rc{(a) Relative degree of superluminosity $\delta u$ as a function of electron density $n_e$ for a plane wave with $a_0 = 10$. (b) Maximum transverse displacement \( y_{\max} \) in the parameter scan shown in \cref{fig:pl_2Dscan}.}}
    \end{center}
\end{figure}

The parameter scan reveals that higher-order resonances impose an additional requirement: the electron must have a larger value of the integral of motion \( S \). As shown in \cref{fig:pl_2Dscan}, \( S \) is directly linked to the initial transverse displacement, which in turn determines the amplitude of transverse oscillations during energy gain. This implies that higher-order resonances require larger betatron amplitudes compared to the betatron resonance (\( \ell = 1 \)). Consequently, the laser beam must be sufficiently wide to support these larger oscillations, setting a more stringent lower bound on the beam width for engaging higher-order resonances.

The parameter scan also shows that higher-order resonances can contribute to generating an energetic electron population even in the low-\( \delta u \) regime. While the betatron resonance is expected to produce the most energetic electrons in this regime, the third-order resonance, for instance, can still generate electrons reaching approximately half that energy — a range that is still of considerable interest. The role of these higher-order resonances in shaping the electron spectrum has yet to be examined in fully self-consistent PIC simulations. Our results suggest that their contribution to the mid-energy range could be significant and warrants closer investigation.

Finally, we note that increasing the plasma density causes the laser phase velocity to become more superluminal, leading to a larger value of \( \delta u \). As a result, higher-order resonances tend to become effective in higher-density regimes where the betatron resonance no longer provides significant energy gain. This can be advantageous in practice, since higher plasma densities are typically more favorable for producing electron populations with greater charge — a benefit for applications that rely on secondary radiation generated by energetic electrons.

\rc{To illustrate the impact of the electron density $n_e$ on $\delta u$, we consider a simple model in which the laser, approximated as a plane wave with $a_0 = 10$, propagates through a uniform plasma. The dispersion relation can be written approximately as $\omega_0^2 \approx k^2 c^2 + \omega_{pe}^2 / a_0$, where $k$ is the wave vector and $\omega_{pe} = \sqrt{4 \pi n_e e^2 / m_e}$ is the plasma frequency. The factor $1/a_0$ accounts for the reduction in effective plasma frequency due to relativistic transparency~\cite{gibbon2004shortRT}. Introducing the critical density $n_{cr}$, defined by $\omega_{pe}^2 = \omega_0^2$, and using $u = \omega_0 / kc$, we obtain from the dispersion relation:
\begin{equation}
    u = \left( 1 - \frac{n_e}{a_0 n_{cr}} \right)^{-1/2}.
\end{equation}
\Cref{fig:vph_ymax}(a) shows the corresponding value of $\delta u = u - 1$ as a function of $n_e / n_{cr}$. According to the scan in \cref{fig:pl_2Dscan}, the peak energy gain for the third-order resonance occurs at $\delta u \approx 1.2 \times 10^{-2}$, which corresponds to a plasma density of $n_e \approx 0.25 n_{cr}$. In comparison, the peak energy gain for the betatron resonance occurs at $\delta u \approx 3 \times 10^{-3}$, requiring a lower density of $n_e \approx 0.06 n_{cr}$.}

\rc{In order to fully benefit from the discussed resonances, the amplitude of the betatron oscillations must remain smaller than the transverse size of both the channel containing the static plasma fields and the laser beam itself. Rather than performing an additional scan that includes these constraints, we use the existing scan to assess the necessary transverse extent. \Cref{fig:vph_ymax}(b) shows the maximum transverse displacement, $y_{\max}/\lambda_0$, from the scan in \cref{fig:pl_2Dscan}. As expected, higher-order resonances lead to greater transverse excursions. For the parameters used in the scan, the betatron resonance requires a channel radius greater than $3 \lambda_0$, while the third-order resonance requires a radius greater than $4 \lambda_0$. It is also important to note that the longitudinal interaction length places an additional requirement. Since the electron gains energy while co-propagating with the laser, the required acceleration distance becomes substantial — particularly for the betatron resonance. In the considered scan, a propagation distance of roughly $2000 \lambda_0$ is needed for the electron to reach its peak energy. While our study does not account for laser depletion or plasma instabilities, these effects may influence feasibility and merit further investigation.}

\appendix

\section{Paraxial approximation} \label{appe: paraxial}

\begin{figure}[htb]
    \begin{center}
    \includegraphics[width=1\columnwidth,clip]{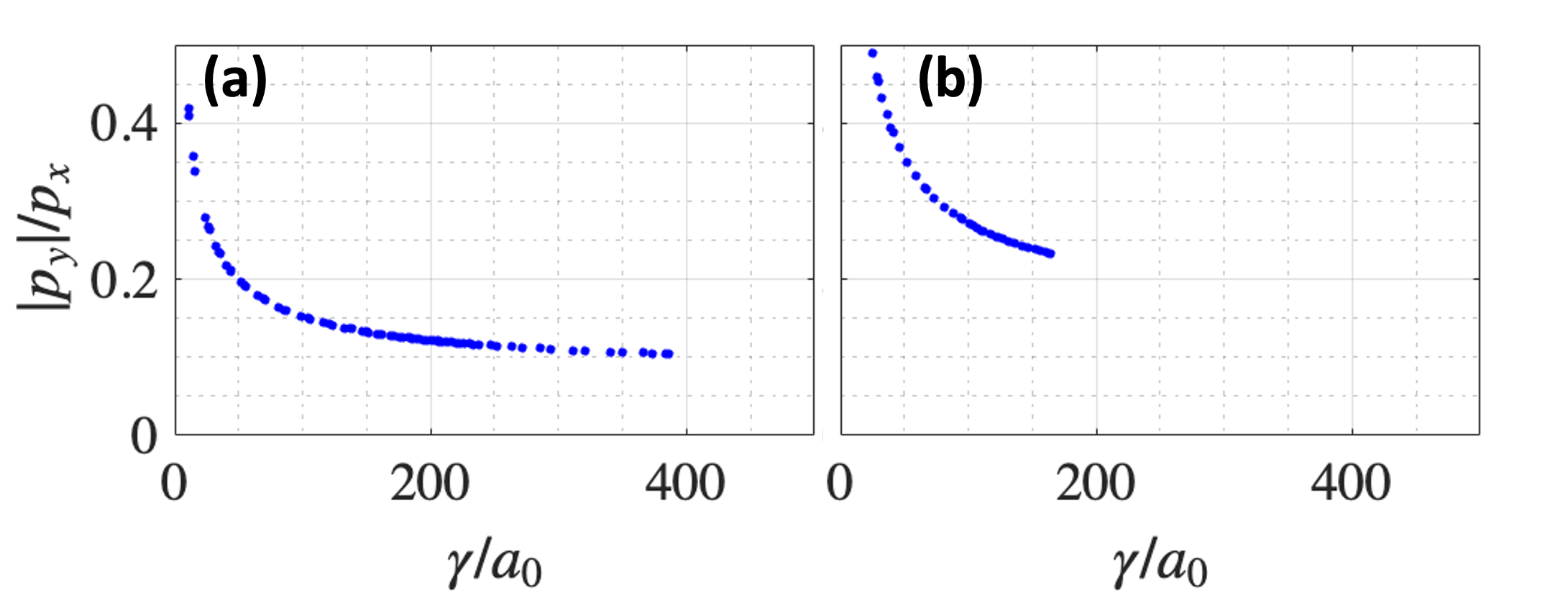}
    \caption{\label{fig:paraxial}  \rc{Ratio \( |p_y|/p_x \) as a function of \( \gamma / a_0 \) for the two examples discussed in the main text: (a) the betatron resonance from \cref{fig:splm_lml}(c) and (b) the third-order resonance from \cref{fig:3rd_res}(a). The ratio is evaluated at instances when the electron crosses the axis (\( y = 0 \)).}} 
    \end{center}
\end{figure}

\rc{\Cref{fig:paraxial} shows the ratio $|p_y|/p_x$ as a function of $\gamma$ for the two examples considered in the main text: the betatron resonance case from \cref{fig:splm_lml}(c) and the third-order resonance case from \cref{fig:3rd_res}(a). The ratio is shown at instances when the electron crosses the axis ($y = 0$), as this is where the ratio reaches its highest value for each betatron oscillation.}

\rc{In both cases, the ratio $|p_y|/p_x$ decreases as the electron gains energy. We conclude from the plots that the condition $p_x \gg |p_y| \gg m_e c$ becomes increasingly well satisfied before the electron reaches its peak energy. In \cref{sec: frequencies}, the analytical expressions for the frequencies are derived assuming that this condition, given by \cref{cond: 1}, is satisfied. The corresponding regime is often referred to as the paraxial approximation. The plots in \cref{fig:paraxial} confirm that the theory of \cref{sec: frequencies} becomes applicable before the electron reaches its peak energy and is well-suited for describing the electron dynamics.}

% To enhance the contrast in the energy gain scan shown in \cref{fig:pl_2Dscan}, we plot the color scale of \( \gamma_{\max}/a_0 \) on a logarithmic scale shown in \cref{fig:scan_log}. This improves the visibility of energy gain features, particularly in the low-\( \delta u \) regime.

% \begin{figure}[t]
%     \begin{center}
%     \includegraphics[width=1\columnwidth,clip]{rebuttal_scan_log.png}
%     \caption{\label{fig:scan_log} \rc{The maximum relativistic factor, \( \gamma_{\max} \), achieved by an electron after propagating a longitudinal distance of \( 2000 \lambda_0 \). The color scale is shown on a logarithmic scale.
% }
% } 
%     \end{center}
% \end{figure} 

\section*{Acknowledgements}

This material is based upon work supported by the Department of Energy National Nuclear Security Administration under Award Number DE-FOA-0004203 and DE-NA0004030. Simulations were performed using HPC resources provided by TACC at the University of Texas and NERSC at Lawrence Berkeley National Laboratory.

\section*{References}
\bibliography{Collection}

%###########################################################################################

\end{document}